\documentstyle[preprint,tighten,aps,amstex,twocolumn,graphicx,floats]{revtex}
\def\ba{\begin{eqnarray}}
\def\ea{\end{eqnarray}}

\begin{document}
\thispagestyle{empty}

\newcommand{\sla}[1]{/\!\!\!#1}

\renewcommand{\small}{\normalsize} 

\preprint{
\font\fortssbx=cmssbx10 scaled \magstep2
\hbox to \hsize{
\hskip.5in \raise.1in\hbox{\fortssbx University of Wisconsin - Madison}
\hfill\vtop{\hbox{\bf MADPH-99-1142}
            \hbox{\bf FERMILAB-Pub-99/290-T}
            \hbox{hepph/9911385}
            \hbox{November 1999}} }
}

\title{\vspace{.45in}
A method for identifying \\
$\mathbf{H \to \tau\tau \to e^{\pm} \mu^{\mp} \sla{p_T}}$ \\
at the CERN LHC
}
\author{T.~Plehn$^1$, D.~Rainwater$^2$ and D.~Zeppenfeld$^1$\\[3mm]}
\address{
$^1$Department of Physics, University of Wisconsin, Madison, WI 53706\\
$^2$Fermi National Accelerator Laboratory, Batavia, IL 60510
}
\maketitle
\begin{abstract}
Weak boson fusion promises to be a copious source of intermediate mass Higgs 
bosons at the LHC. The additional very energetic forward jets in these events 
provide for powerful background suppression tools. We analyze the subsequent
$H\to\tau\tau \to e^{\pm} \mu^{\mp} \sla{p_T}$ decay for Higgs boson masses in 
the 100-150~GeV range. A parton level analysis of the dominant backgrounds 
demonstrates that this channel allows the observation of $H\to\tau\tau$ in a 
low-background environment, yielding a significant Higgs boson signal with an 
integrated luminosity of order 60~fb$^{-1}$ or less, over most of the mass 
range. We also restate a No-Lose Theorem for observation of at least one of the 
CP-even neutral Higgs bosons in the MSSM, which requires an integrated 
luminosity of only 40~fb$^{-1}$.
\end{abstract}

\vspace{0.2in}


\section{Introduction}
\label{sec:intro}

The search for the Higgs boson and, hence, for the origin of electroweak 
symmetry breaking and fermion mass generation, remains one of the premier 
tasks of present and future high energy physics experiments. Fits to 
precision electroweak (EW) data have for some time suggested a relatively 
small Higgs boson mass, of order 100~GeV~\cite{EWfits}, which is also the 
preferred mass range for the lightest Higgs boson in the minimal 
supersymmetric extension of the Standard Model (MSSM). 

For the intermediate mass range, most of the literature has focused on Higgs 
boson production via gluon fusion~\cite{reviews} and $t\bar{t}H$~\cite{ttH} or 
$WH(ZH)$~\cite{WH} associated production. Cross sections for Standard Model 
(SM) Higgs boson production at the LHC are well-known~\cite{reviews}, and while 
production via gluon fusion has the largest cross section by almost one order 
of magnitude, there are substantial QCD backgrounds. A search for the very 
clean four-lepton signature from $H\to ZZ$ decay can find a Higgs boson in the 
mass region $M_H \gtrsim 125$~GeV, but due to the small branching fraction of 
this mode very large integrated luminosities, up to 100~fb~$^{-1}$ or more, 
are required. For Higgs boson masses less than about 140 GeV, the inclusive 
search for $H\to\gamma\gamma$ events is usually considered the most promising 
strategy~\cite{CMS-ATLAS}, while for $140 < M_H < 200$~GeV the most promising 
search is for decay to $W$ pairs~\cite{CMS-ATLAS,RZ_WW,thesis,DittDrein}.

The search for MSSM Higgs bosons must include neutral CP even and CP odd
mass eigenstates, as well as charged ones. The upper mass limit of 
$\sim$~130~GeV~\cite{one_loop,two_loop} on the light scalar makes it 
look similar to its intermediate-mass Standard Model analogue, for 
large regions of the MSSM parameter space. While 
one would expect the most promising channel to again be $\gamma\gamma$ 
decay~\cite{reviews,CMS-ATLAS}, the branching ratio for this mode 
is even smaller than in the Standard Model.

The second largest production cross section is predicted for weak-boson 
fusion (WBF), $qq \to qqVV \to qqH$. These events contain additional 
information in their observable quark jets. Techniques like forward jet 
tagging~\cite{Cahn,BCHP,DGOV} can then be exploited to significantly reduce 
the backgrounds. WBF and gluon fusion nicely complement each other: 
together they allow for a measurement of the $t\bar{t}H/WWH$ coupling ratio.

Another feature of the WBF signal is the lack of color exchange between the 
initial-state quarks. Color coherence between initial- and final-state gluon 
bremsstrahlung leads to suppressed hadron production in the central region, 
between the two tagging-jet candidates of the signal~\cite{bjgap}. This is in 
contrast to most background processes, which typically involve color exchange 
in the $t$-channel and thus lead to enhanced hadronic activity between the 
tagging jets. We exploit these features, via a veto of soft jet activity 
in the central region~\cite{bpz_mj}.

We have previously established the feasibility of WBF intermediate-mass Higgs 
production as both a discovery channel (via 
$H\to W^{(*)}W^{(*)} \to e^\pm \mu^\mp \sla{p_T}$ 
decays~\cite{RZ_WW,thesis}) and as a means to provide the first direct 
Higgs boson-fermion coupling measurement~\cite{RZ_tautau,thesis} 
($H\to\tau\tau\to h^\pm \ell^\mp \sla{p_T}$).
The latter allows one to na\"{\i}vely extend the Standard Model search to 
the MSSM case: the structure of the Higgs sector predicts at least one 
scalar in the intermediate mass range, rendering the $\tau\tau$ channel a 
crucial test of the MSSM~\cite{PRZ_mssm}. 
Here, we show how the additional channel 
$H\to\tau\tau\to e^\pm \mu^\mp \sla{p_T}$ can be isolated, effectively 
doubling the available statistics for a measurement of the $H\tau\tau$ 
coupling.

Our analysis is a parton-level Monte Carlo study, using full tree-level 
matrix elements for the WBF Higgs signal and the various backgrounds. 
In Section~\ref{sec:calc} we describe our calculational tools, the 
methods employed in the simulation of the various processes, and important 
parameters. In Section~\ref{sec:analysis} 
we demonstrate how forward jet tagging, a $b$ veto, and lepton cuts can be 
combined to yield an $\approx \;$2/1 to 1/2 signal-to-background (S/B) ratio, 
depending on the Higgs mass. 
The different minijet patterns in signal and background processes are 
discussed in Section~\ref{sec:minijet}. We describe how they can be used to 
achieve additional large suppression of the QCD backgrounds relative to 
the signal. Combined with the results of Section~\ref{sec:analysis} this 
yields production cross sections of signal and backgrounds as given in 
Table~\ref{summary}, which summarizes our results. 
In Section~\ref{sec:MSSM} we reanalyze the impact
on covering the MSSM parameter space and discuss the luminosity 
requirement at the LHC for which the entire $m_A,\tan \beta$ plane can be 
covered. A final discussion of our results and conclusions is given in 
Section~\ref{sec:disc}.


\section{Calculational Tools}
\label{sec:calc}

We simulate $pp$ collisions at the CERN LHC, $\protect\sqrt{s} = 14$~TeV. 
All signal and background cross sections are determined in terms of full 
tree-level matrix elements for the contributing subprocesses and are 
discussed in more detail below.

For all our numerical results we have chosen $1/\alpha = 128.933$, 
$M_Z = 91.187$~GeV, and $G_F = 1.16639\cdot 10^{-5}\;{\rm GeV}^{-2}$, which 
translates into $M_W = 79.963$~GeV and ${\rm sin}^2\theta_W = 0.2310$ when 
using the tree-level relations between these input parameters. This value for 
$M_W$ is somewhat lower than the current world average of $80.39$ GeV.
However, this difference has negligible effects on all cross sections, e.g. 
the $qq\to qqH$ signal cross section varies by about $0.5\%$ between these two 
$W$ mass values. The tree level relations between the input parameters are 
kept in order to guarantee electroweak gauge invariance of all amplitudes. 
For all QCD effects, the running of the strong coupling constant is 
evaluated at one-loop order, with $\alpha_s(M_Z) = 0.118$. We employ 
CTEQ4L parton distribution functions~\cite{CTEQ4_pdf} throughout. Unless 
otherwise noted the factorization scale is chosen as $\mu_f =$ min($p_T$) 
of the defined jets.


\subsection{The $\mathbf{qq\to qqH(g)}$ signal process (and background)}
\label{sec:qqH}

The signal can be described, at lowest order, by two single-Feynman-diagram
processes, $qq \to qq(WW,ZZ) \to qqH$, i.e. $WW$ and $ZZ$ fusion where 
the weak bosons are emitted from the incoming quarks~\cite{qqHorig}. 
Because of the small Higgs boson width in the mass range of interest, these 
events can reliably be simulated in the narrow width approximation. 
From previous studies of $H\to\gamma\gamma$~\cite{RZ_gamgam}, 
$H\to\tau\tau$~\cite{RZ_tautau} and $H\to WW$~\cite{RZ_WW} decays in weak 
boson fusion we know several features of the signal, which can be exploited 
here also: the centrally produced Higgs boson tends to yield central decay 
products (in this case $\tau^+\tau^-$), and the two quarks enter the 
detector at large rapidity compared to the $\tau$'s and with transverse 
momenta in the 20 to 100 GeV range, thus leading to two observable forward 
tagging jets.

For the study of a central jet veto, we utilize the results of previous 
studies where we simulated the emission of at least one extra 
parton~\cite{thesis,RZ_tautau,RSZ_vnj}. This was achieved by calculating 
the cross sections for the process $qq\to qqHg$, i.e. weak boson fusion 
with radiation of an additional gluon, and all crossing related processes. 

We note that the signal simulations, with decays to tau pairs replaced by 
decays to $W$ pairs, which in turn decay leptonically, will ultimately also 
be a source of background for the $H\to\tau\tau$ signal under study.


\subsection{The QCD $\mathbf{t \bar{t} + jets}$ backgrounds}
\label{sec:ttnj}

Given the H decay signature, the main physics background to our 
$e^{\pm}\mu^{\mp}\sla p_T$ signal arises from $t\bar{t} + jets$ production, 
due to the large top production cross section at the LHC and because 
the branching ratio $B(t\to Wb)$ is essentially $100\%$. 

The basic process we consider is $pp\to t\bar{t}$, which can be either $gg$- 
or $q\bar{q}$-initiated, with the former strongly dominating at the LHC. 
QCD corrections to this lead to additional real parton emission, i.e. 
to $t\bar{t} + j$ events. Relevant subprocesses are 
\ba
g q  \to t \bar{t} q \, , \; \; g \bar{q} \to t \bar{t} \bar{q} \, , \; \;
q \bar{q} \to t \bar{t} g \, , \; \; g g \to t \bar{t} g \, , 
\nonumber
\ea
and the subprocesses for $t\bar{t} + jj$ events can be obtained similarly. 
For the case of no additional partons, the $b$'s from the decaying top quarks 
may be identified as the tagging jets. At the same time, we can identify a 
distinctly different, perturbative region of phase space, where the 
final-state light quark or gluon gives rise to one tagging jet, and one of 
the two decay $b$'s is identified as the other tagging jet. Finally, there 
is a third distinct region of phase space, for the perturbative hard process 
$pp\to t\bar{t} + jj$, where the final-state light quarks or gluons are the 
two tagging jets. The $t\bar{t}$ and $t\bar{t}j$ matrix elements were 
constructed using {\sc madgraph}~\cite{Madgraph}, while the $t\bar{t}jj$ 
matrix elements are from Ref.~\cite{Stange}.

Decays of the top quarks and $W$'s are included in the matrix elements; 
however, while the $W$'s are allowed to be off-shell, the top quarks are 
required to be on-shell. This approximation neglects the contribution
from $Wt$ production, which has been shown to be comparable to $t\bar t$
rates in studies of the $H\to WW$ signal~\cite{DittDrein,CMS-ATLAS}. We will
compensate by being conservative in assessing minijet veto probabilities
for top backgrounds. Note that these approximations are not critical because
backgrounds with real $W$ pairs can be distinguished quite effectively from
$\tau\tau$ events and the real top decay backgrounds that we consider will 
be shown to constitute a minor fraction of the final backgrounds.  
In the calculation of the $t\bar t$ background 
energy loss from $b\to\ell\nu X$ is included to 
generate more accurate $\sla{p}_T$ distributions. In all cases, the 
factorization scale is chosen as $\mu_f =$ min($E_T$) of the massless 
partons/top quarks. As in our earlier work~\cite{RZ_WW}, the overall 
strong coupling constant factors are taken as 
$(\alpha_s)^n = \prod_{i=1}^n \alpha_s(E_{T_i})$, where the product runs 
over all light quarks, gluons and top quarks.


\subsection{The QCD $\mathbf{b \bar{b} + jj}$ background}
\label{sec:bbjj}

The semileptonic decays of bottom or charm quarks provide another source 
of leptons and neutrinos which can be misidentified as tau decays. These 
heavy quark pairs are produced strongly and a priori one is dealing with 
a very large potential background. It can be reduced by several 
orders of magnitude, however, by requiring the leptons from the decay of 
the heavy quarks to be isolated. Because of the softer fragmentation function
of a $c$-quark as compared to $b$-quarks, leptons from charm decay are much
less likely to be isolated than $b$-decay leptons. In 
the phase space region of interest to us, where both heavy quarks must reside 
in the central angular region and have substantial transverse momentum, 
the production cross sections for charm and bottom pairs are roughly equal. 
As a result we consider only the $b$-quark background in the following. 

In addition to the two high transverse momentum $b$-quarks, which both 
must undergo semileptonic decay, two forward tagging jets will be
required as part of the signal event selection. The relevant leading
order process therefore is the
production of $b\bar{b}$ pairs in association with two jets, which
includes the subprocesses
\ba
    gg          \rightarrow b\bar{b} gg \; , \; \;
    qg          \rightarrow b\bar{b} qg \; , \; \;
    q_{1} q_{2} \rightarrow b\bar{b} q_{1} q_{2} \, . \nonumber
\ea
The exact matrix elements for the ${\cal O} (\alpha_{s}^{4})$ processes
are evaluated, including all the crossing related subprocesses, and retaining
a finite $b$-quark mass~\cite{Stange}.
The Pauli interference terms between identical quark flavors in the process
$q_{1}q_{2}\rightarrow b\bar{b}q_{1}q_{2}$ are neglected, with little effect
in the overall cross section, due to the large difference in the rapidity
of the final state light quarks. The factorization and renormalization 
scales are chosen as in the analogous $t\bar tjj$ case.

The semileptonic decay $b\to\nu\ell c$ of both of the $b$-quarks is simulated
by multiplying the $b\bar{b}jj$ cross section by a branching ratio factor of
0.0218 (corresponding to an $e^+\mu^-$ or $\mu^+e^-$ final state) and by
implementing the $V-A$ decay distributions of the $b$-quarks in the collinear
limit. The collinear approximation for the $b\to\nu\ell c$ decay is appropriate
here because the lepton transverse momentum and $\sla{p}_T$ cuts to be 
imposed below force the parent $b$-quarks to move relativistically in the lab.
Denoting the neutrino and charged lepton energy fractions by $x_\nu$ and
$y_\ell$, respectively, the double differential $b$-quark decay distribution 
is given by~\cite{bclnu}
\begin{align}
\label{eq:d2Gdxdy}
{1\over \Gamma} {d^2\Gamma\over dx_\nu dy_\ell} = {2c\over f(r)} \;
\biggl( & c\;(1-x_\nu)\;\left[ c+(3-c)\;x_\nu\; \right] \nonumber\\
        & + 3ry_\ell\;{(2-c)\;x_\nu+c \over 1-x_\nu-y_\ell} \; \biggr)
\end{align}
assuming an unpolarized initial $b$-quark. Here $r = m_c^2/m_b^2$,
and the dependence on the final state charm quark mass, $m_c$, is absorbed 
into the correction term
\begin{equation}
c = {1-r-x_\nu-y_\ell\over 1-x_\nu-y_\ell} = 1-{r\over z_c}\; .
\end{equation}
Finally, $f(r)$ is the width suppression factor for the $b\to \nu\ell c$
decay due to the finite charm quark mass,
\begin{equation}
\label{eq:width}
f(r) = (1-r^2)(1-8r+r^2) - 12r^2\log r \; .
\end{equation}
In our numerical simulations we set $m_b = 5.28$~GeV and $m_c=1.87$~GeV, 
i.e. we use the lightest meson masses in order to approximately obtain 
the correct kinematics for the heavy quark decays. 
In Ref.~\cite{Cavalli} a factor 100 reduction of the $b\bar{b}$ background
was found as a result of lepton isolation for a single $b\to \nu\ell c$ decay,
requiring $E_T<5$~GeV in a cone
of radius 0.6 around the charged lepton of $p_{T\ell}>20$~GeV. 
In our simulation, after energy
smearing of the charm quark jet (see below), we reproduce this reduction 
factor. The suppression from lepton isolation is smaller for lower
$p_{T\ell}$ cuts. We model these effects by using Eq.~(\ref{eq:d2Gdxdy}).

Since our suppression of $b\to \nu\ell c$ decays from lepton isolation
strongly depends on the energy resolution assumed for the very soft charm 
quark jet, the determination of heavy quark backgrounds should eventually 
be repeated with a full detector simulation. We will show, however, that the
$b\bar{b}jj$ background is truly negligible after all the selection cuts 
to be described in this paper. Therefore, the approximate treatment of these 
backgrounds is sufficient for our purposes.


\subsection{The QCD and EW $\mathbf{\tau^+\tau^- +jj}$ backgrounds}
\label{sec:tau}

The next obvious backgrounds arise from Z decays to real $\tau$'s which then 
decay leptonically. Thus, we need to study real-emission QCD corrections to 
the Drell-Yan process $q\bar{q} \to (Z,\gamma) \to \tau^+\tau^-$. 
For $\tau^+ \tau^- jj$ events these background processes include~\cite{Kst}
\ba
q g \to q g \tau^+ \tau^- , \qquad  q q' \to q q' \tau^+ \tau^- ,
\nonumber
\ea
which are dominated by $t$-channel gluon exchange, and all crossing-related 
processes, such as
\ba
q \bar{q} \to g g \tau^+ \tau^- , \qquad g g \to q \bar{q} \tau^+ \tau^- .
\nonumber
\ea
All interference effects between virtual photon and $Z$-exchange are included.
We call these processes collectively the ``QCD $\tau\tau jj$'' background. 
Similar to the treatment of the signal processes, we use a parton-level 
Monte-Carlo program based on the work of Ref.~\cite{BHOZ} to model the QCD 
$\tau\tau jj$ background.

From our study of $H\to\tau\tau$ in weak boson fusion~\cite{RZ_tautau}, we 
know that the EW (t-channel weak boson exchange) cross section for 
$Zjj$ production will be 
comparable to the QCD cross section in the phase space region of interest. 
We use the results of Ref.~\cite{CZ_gap} for modeling 
the EW $\tau\tau jj$ background.

The dual leptonic decays of the $\tau$'s are simulated by multiplying the 
$\tau^+\tau^-jj$ cross section by a branching ratio factor of $(0.3518)^2/2$ 
and by implementing the lepton energy distributions for collinear tau decays, 
with helicity correlations included 
as in our previous analysis of $H\to\tau\tau$~\cite{RZ_tautau}.


\subsection{The QCD $\mathbf{WW+jj}$ background}

We must further consider any other significant source of one electron, one 
muon and significant $\sla{p}_T$ to make a realistic analysis of the 
backgrounds. 
An obvious candidate arises from real-emission QCD corrections to $W^+W^-$ 
production, with subsequent decay of the two $W$`s to electrons or muons.  
For $W^+ W^- jj$ events these background processes include~\cite{VVjj}
\ba
q g \to q g W^+ W^- , \qquad  q q' \to q q' W^+ W^- ,
\nonumber
\ea
which are dominated by $t$-channel gluon exchange, and all crossing
related processes, such as
\ba
q \bar{q} \to g g W^+ W^- , \qquad g g \to q \bar{q} W^+ W^- .
\nonumber
\ea
We call these processes collectively the ``QCD $WWjj$'' background. To 
estimate the minijet activity in these events we use the results 
for QCD $Z + jets$ processes, which are kinematically 
similar~\cite{RZ_WW,thesis}.

Note that we neglect $W \to\tau\nu \to\ell\nu\nu$ decays in our simulation of 
$WWjj$ backgrounds. This is justified by the suppressed
leptonic branching ratio of the $\tau$ decays. We show below that the 
$WW\to e\mu\nu\nu$ backgrounds are already negligible and, therefore, 
the extra $W\to\tau\nu$ decays do not need to be analyzed in detail.


\subsection{The EW $\mathbf{WW+jj}$ background}

These backgrounds, analogous to QCD $WWjj$ production, arise from $W^+W^-$ 
bremsstrahlung in quark-(anti)quark scattering via $t$-channel electroweak 
boson exchange, with subsequent decay $W^+W^-\to\ell^+\ell^-\sla p_T$:
\ba
qq' \to qq' W^+W^-
\nonumber
\ea
Na\"{\i}vely, this EW background may be thought of as suppressed compared 
to the analogous QCD process above. However, it includes 
electroweak boson fusion, $VV \to W^+W^-$ via $s$- or $t$-channel 
$\gamma/Z$-exchange or via $VVVV$ 4-point vertices, which has a momentum 
and color structure identical to the signal. Thus, it cannot easily be 
suppressed via cuts.

The matrix elements for these processes were constructed using 
{\sc madgraph}~\cite{Madgraph}. We include charged-current (CC) and 
neutral-current (NC) processes, but discard s-channel EW boson and t-channel 
quark exchange processes as their contribution was found to be only 
$\approx 1\%$, while adding significantly to the CPU time needed for the 
calculation. In general, for the regions of phase space containing 
far-forward and -backward tagging jets, s-channel processes are severely 
suppressed. We refer collectively to these processes as the ``EW $WWjj$'' 
background. Both $W$'s are allowed to be off-shell, and all off-resonance 
graphs are included. In addition, the Higgs boson graphs must be included 
to make the calculation well-behaved at large $W$-pair invariant masses. 
However, it is convenient to separate continuum $W$-pair production from the 
very narrow $H\to W^+W^-$ resonance. We do this by setting $M_H$ to 60~GeV 
in the EW $WWjj$ background which effectively removes the $s$-channel Higgs 
contribution. The $H\to W^+W^-$ background is then calculated separately for `
each Higgs boson mass under consideration.  
A clean separation of the Higgs boson signal and the EW $WWjj$ background is 
possible because interference effects between the two are negligible for the 
Higgs boson mass range of interest.

The effects of additional gluon radiation are estimated by using the
results of Refs.~\cite{RZ_WW,thesis} for EW  $\tau\tau jj$ events, 
which are directly applied here. The EW $\tau\tau jj$ and EW $WWjj$ 
backgrounds are quite similar kinematically, which justifies the use of the 
same veto probabilities for central jets. 


\subsection{Detector resolution}

The QCD processes discussed above lead to steeply falling jet transverse 
momentum distributions. As a result, finite detector resolution can have a 
sizable effect on cross sections. These resolution effects are taken into 
account via Gaussian smearing of the energies of jets/$b$'s and charged 
leptons.  We use 
\begin{equation}
{\triangle{E} \over E} =
{3.3 \over E} \oplus {0.6 \over {\sqrt E}} \oplus 0.03 \; ,
\end{equation}
for central jets (with individual terms added in quadrature), based on ATLAS 
expectations~\cite{CMS-ATLAS}. For charged leptons we 
use
\begin{equation}
{\triangle{E} \over E} = 2\% \; .
\end{equation}

In addition, finite detector resolution leads to fake 
missing-transverse-momentum in events with hard jets. An ATLAS 
analysis~\cite{Cavalli} showed 
that these effects are well parameterized by a Gaussian distribution of 
the components of the fake missing transverse momentum vector, 
$\vec\sla p_T$, with resolution 
\begin{equation}
\sigma(\sla p_x,\sla p_y) = 0.46 \cdot \sqrt{\sum{E_{T,had}}} \; ,
\end{equation}
for each component. In our calculations, these fake missing transverse 
momentum vectors are added linearly to the neutrino momenta.


\section{Higgs signal and backgrounds}
\label{sec:analysis}

The $qq\to qqH, \; H\to\tau\tau\to e^\pm \mu^\mp \nu \bar{\nu}$ double
leptonic decay signal is characterized by two forward jets and the $\tau$ 
decay leptons ($e,\mu$). Before discussing background levels and further 
details like minijet radiation patterns, we need to identify the search region 
for these hard $Hjj$ events. The task is identical to the Higgs searches in 
$qq\to qqH,\;H\to\gamma\gamma,\tau\tau,WW$ which were considered 
previously~\cite{RZ_WW,thesis,RZ_tautau,RZ_gamgam}. We can thus adopt the 
strategy of these earlier analyses and start out by discussing a basic level 
of cuts on the $qq\to qqH,\;H\to\tau\tau$ signal. Throughout this section we 
assume a Higgs mass of $M_H = 120$~GeV for illustration purposes, 
but we do not optimize cuts for this mass.

The minimum acceptance requirements ensure that the two jets and two charged 
leptons are observed inside the detector (within the hadronic and 
electromagnetic calorimeters, respectively), and are well-separated from each 
other:
\ba
\label{eq:basic}
& p_{T_j} \geq 20~{\rm GeV} \, , \; \; |\eta_j| \leq 5.0 \, , \; \;
\triangle R_{jj} \geq 0.7 \, , \nonumber\\
& p_{T_\ell} \geq 10~{\rm GeV} \, ,\; \;
|\eta_{\ell}| \leq 2.5 \, , \; \; \triangle R_{j\ell} \geq 0.7 \, .
\ea
The charged leptons must be isolated in order to reduce backgrounds from 
heavy quark decays. Thus a minimum angular distance must be imposed on the
electron and the muon signaling the tau decays: 
\begin{equation}
\label{eq:basic+}
\triangle R_{e\mu} \geq 0.4 \, .
\end{equation}
This has negligible effect on the Higgs boson signal.

A feature of the QCD $Zjj$ and $WWjj$ backgrounds is the generally higher
rapidity of the $Z$ or $W$'s as compared to the Higgs signal: weak boson 
bremsstrahlung occurs at small angles with respect to the parent quarks,
producing a $Z$ or $W$'s forward of the jets. Thus, we also require both 
$\ell$'s to lie between the jets with a separation in pseudorapidity 
$\triangle \eta_{j,\ell} > 0.7$, and the jets to occupy opposite hemispheres:
\begin{align}
\label{eq:lepcen}
& \eta_{j,min} + 0.7 < \eta_{\ell_{1,2}} < \eta_{j,max} - 0.7 \, , \notag \\
& \eta_{j_1} \cdot \eta_{j_2} < 0
\end{align}
Finally, to reach the starting point for our consideration of the signal and 
various backgrounds, a wide separation in pseudorapidity is required between 
the two forward tagging jets,
\begin{equation}
\label{eq:gap}
\triangle \eta_{tags} = |\eta_{j_1}-\eta_{j_2}| \geq 4.4 \, ,
\end{equation}
leaving a gap of at least 3 units of pseudorapidity in which the charged 
leptons can be observed. Forward jet tagging has been discussed as an
effective technique to separate weak boson scattering from various 
backgrounds in the past~
\cite{Cahn,BCHP,DGOV,bjgap,bpz_mj,RZ_tautau,RZ_WW,thesis,RZ_gamgam,RSZ_vnj},
in particular for heavy Higgs boson searches. Line 1 of 
Table~\ref{taudata_all} shows the effect of the above cuts on the signal 
and backgrounds for a SM Higgs 
boson of mass $M_H = 120$~GeV. Overall, about $13\%$ of all 
$H\to\tau\tau \to e^\pm \mu^\mp \nu \bar{\nu}$ events generated in weak 
boson fusion are accepted by the cuts of Eqs.~(\ref{eq:basic})-(\ref{eq:gap}) 
(for $M_H = 120$~GeV). 

As is readily seen from the first line of Table~\ref{taudata_all}, the 
dominant backgrounds are $e,\mu$ pairs from heavy quark decays. Of the 
$t\bar{t} (+ jets)$ events 14~fb are from $t\bar{t}$, $360$~fb are 
from $t\bar{t}j$, and the remaining $860$~fb arise from $t\bar{t}jj$ 
production. The additional jets (corresponding to massless partons) are 
required to be identified as far forward tagging jets. The $t\bar{t}jj$ 
cross section is largest because the $t\bar{t}$ pair is not forced to have 
as large an invariant mass as in the first two cases, where one or both $b$'s 
from the decay of the top quarks must pass the tagging jet cuts. 

For the events where only one or none of the $b$'s are identified as a
forward jet, the $b$'s will most frequently lie between the two tagging jets, 
in the region where we search for the $W$ decay leptons. Vetoing events with 
these additional $b$ jets provides a powerful suppression tool to control the 
top background~\cite{RZ_WW}. Note that this does {\it not} require a $b$-tag, 
merely rejection of any events that have an additional jet, which in this 
case would be from a $b$-quark and its decay products. 
(It is quite possible that b-tagging could 
improve this simple rejection criterion, especially in the $p_T < 20$~GeV 
region.) We discard all events where 
a $b$ or $\bar{b}$ with $p_T > 20$~GeV is located in the gap region 
between the tagging jets,
\begin{equation}
\label{eq:bveto}
p_{T_b} > 20\;{\rm GeV} \, , \; \;
\eta_{j,min} < \eta_{b} < \eta_{j,max} \, .
\end{equation}

\begin{figure}[tb]
\includegraphics[width=6.3cm,angle=90]{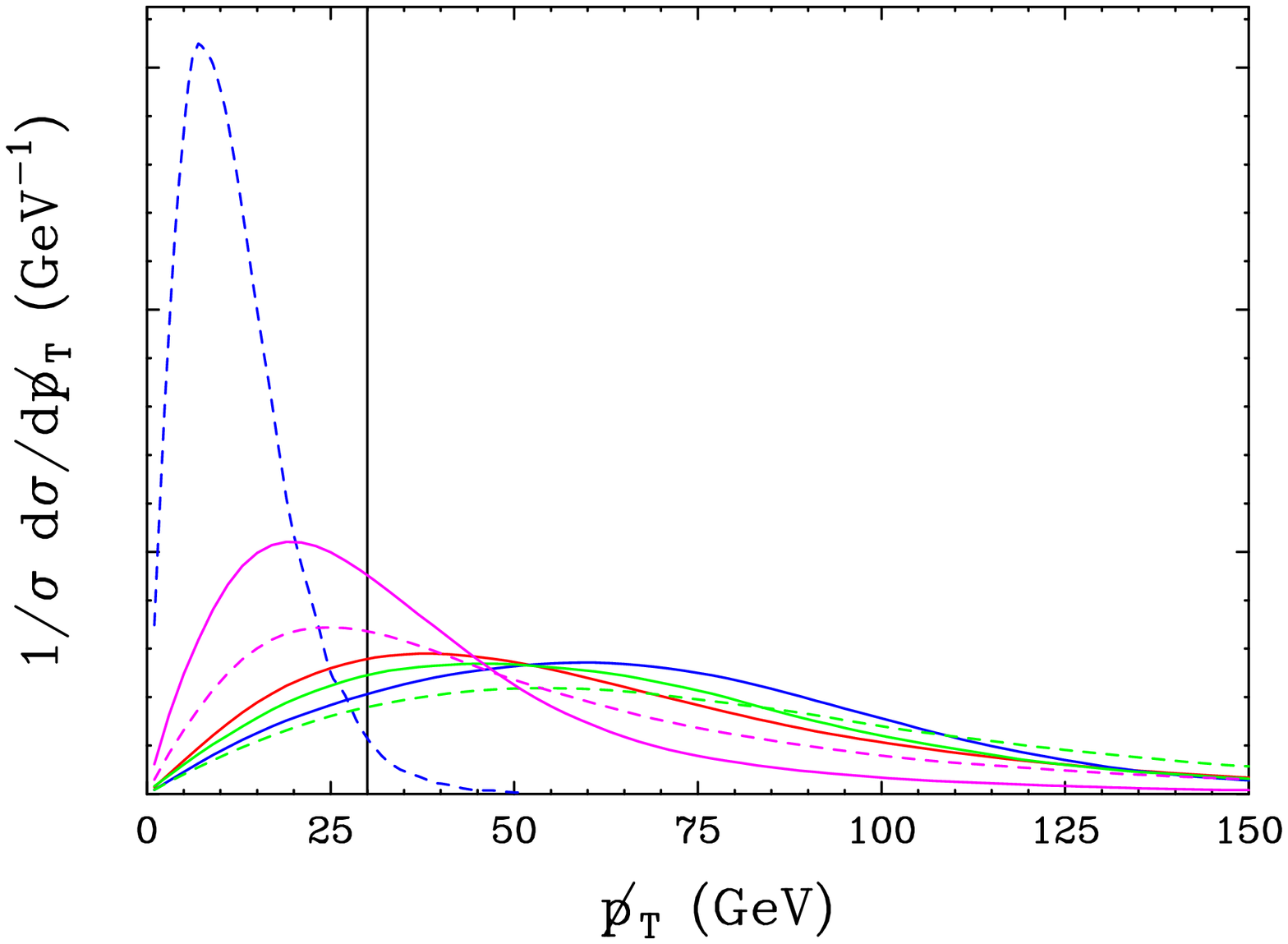} \\
\includegraphics[width=6.3cm,angle=90]{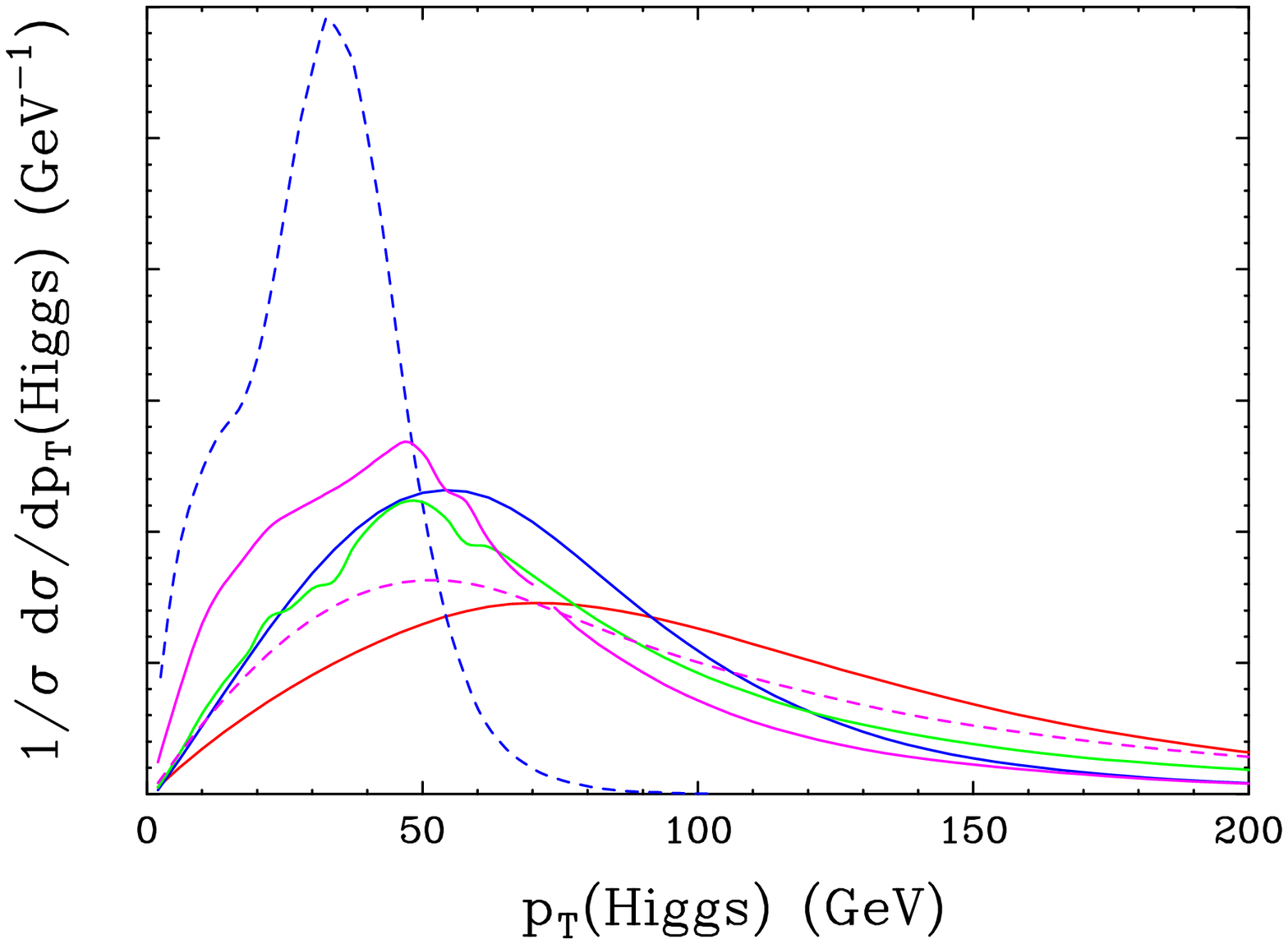} \vspace*{0.2cm}
\caption{Upper: Normalized $\sla{p}_T$ distribution for the signal (red) 
and various backgrounds: 
$t\bar{t} + jets$ (solid blue), $b\bar{b}jj$ (dashed blue), 
QCD $WWjj$ (solid green), EW $WWjj$ (dashed green),
QCD $\tau\tau jj$ (solid magenta) and 
EW $\tau\tau jj$ (dashed magenta). The cuts of 
Eqs.~(\ref{eq:basic})-(\ref{eq:bveto}) are imposed.
Lower: The same for the normalized $p_T$ distribution of the 
reconstructed Higgs boson, except that QCD and EW $WWjj$ contributions 
have been combined (solid green).}
\label{fig:ptmiss}
\label{fig:ptH}
\end{figure}

\begin{figure}[tb]
\includegraphics[width=6.4cm,angle=90]{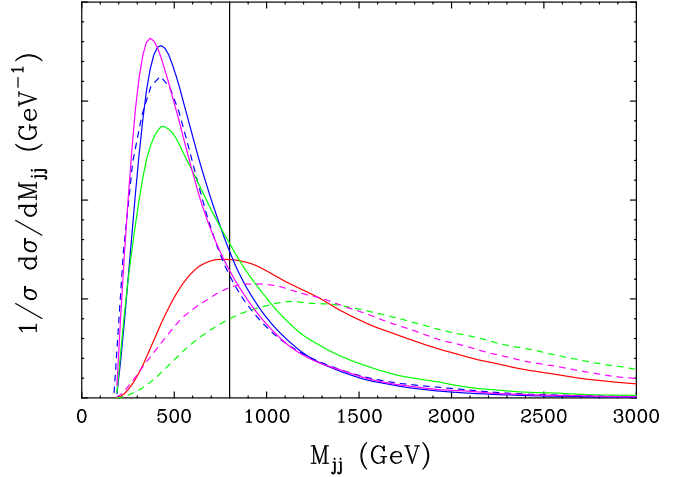} \vspace*{0.2cm}
\caption{Normalized invariant mass distribution of the two tagging jets 
for the signal and the various backgrounds as in 
Fig.~\protect\ref{fig:ptmiss}. The cuts of 
Eqs.~(\ref{eq:basic})-(\ref{eq:bveto}) are imposed. (The distributions 
are essentially unchanged after imposing the additional cut of 
Eq.~\ref{eq:ptmiss}.)}
\label{fig:Mjj}
\end{figure}

This leads to a reduction of $t\bar{t}j$ events by a factor 7 while 
$t\bar{t}jj$ events are suppressed by a factor 100, resulting in cross 
sections of 50 and 8.7~fb, respectively, at the level of the forward 
tagging cuts of Eqs.~(\ref{eq:basic})-(\ref{eq:gap}), which are now 
comparable to the irreducible backgrounds, real taus from $Zjj$ events. 
(See the second line of Table~\ref{taudata_all}.) 
Note that the much higher $b$ veto probability for $t\bar{t}jj$ events results 
in a lower cross section than that for $t\bar{t}j$ events, an ordering which 
will remain even after final cuts have been imposed (see below).

The large $b\bar{b}jj$ background is most effectively reduced by requiring 
a significant level of missing transverse momentum in the event. The 
$\sla{p}_T$ distributions for the signal and the various backgrounds are 
shown in Fig.~\ref{fig:ptmiss}. The extremely soft $\sla{p}_T$ distribution 
for $b\bar{b}$ events is mostly due to the stringent lepton isolation 
requirements. A low transverse momentum of the charm quark in the 
$b\to c\ell\nu$ decay requires a fairly soft parent $b$-quark, which in turn
does not permit a large $\sla{p}_T$ to be carried away by the escaping 
neutrino. This effect is amplified by the QCD nature of the
$b\bar{b}jj$ background which favors the production of low $p_T$ $b$-quarks
in the first place. All other backgrounds involve the decay of one or more 
massive objects (a $Z$, $W$s, or top quarks) into leptons and neutrinos and 
thus result in a much harder $\sla{p}_T$ distribution. The distributions
of Fig.~\ref{fig:ptmiss} motivate a cut 
\begin{equation}
\label{eq:ptmiss}
\sla{p}_T \; > \; 30~{\rm GeV} \; ,
\end{equation}
which brings the $b\bar{b}jj$ background to a manageable level. The
cross sections after this $\sla{p}_T$ cut are shown in the third line 
of Table~\ref{taudata_all}.

A similar reduction of the $b\bar{b}jj$ background can be achieved by a
harder lepton transverse momentum requirement than the 10~GeV cut of
Eq.~(\ref{eq:basic}). However, the signal distribution is quite soft as
well and a harder $p_{T_\ell}$-cut would lead to an undesirable loss of
signal rate. Another option is to make use of the transverse momentum
of the Higgs boson candidate, defined as the recoil needed to balance the 
transverse momentum of the observed hadrons in the event. These 
``Higgs boson'' transverse momentum distributions are also plotted 
in Fig.~\ref{fig:ptH}. They are qualitatively similar to 
$\sla{p}_T$ for all processes, but the peak is shifted to lower values than 
that of the real Higgs boson signal for all backgrounds. While we do not 
use this distribution here, we point out that it may be useful once a 
multivariate analysis is performed at the detector level.

QCD processes at hadron colliders typically occur at smaller invariant 
masses than EW processes, due to the dominance of gluons at small Feynman 
$x$ in the incoming protons. We observe this behavior here, as shown in 
Fig.~\ref{fig:Mjj}. 
The three $t\bar{t} + jets$ backgrounds have been combined for clarity, 
even though their individual distributions are slightly different. One can 
significantly reduce the QCD backgrounds by imposing a lower bound 
on the invariant mass of the two tagging jets: 
\begin{equation}
\label{eq:mjj}
M_{jj} > 800~{\rm GeV} \; .
\end{equation}
Resulting cross sections are shown in the fourth line of 
Table~\ref{taudata_all}.

\begin{figure}[tb]
\includegraphics[width=8.0cm,angle=90]{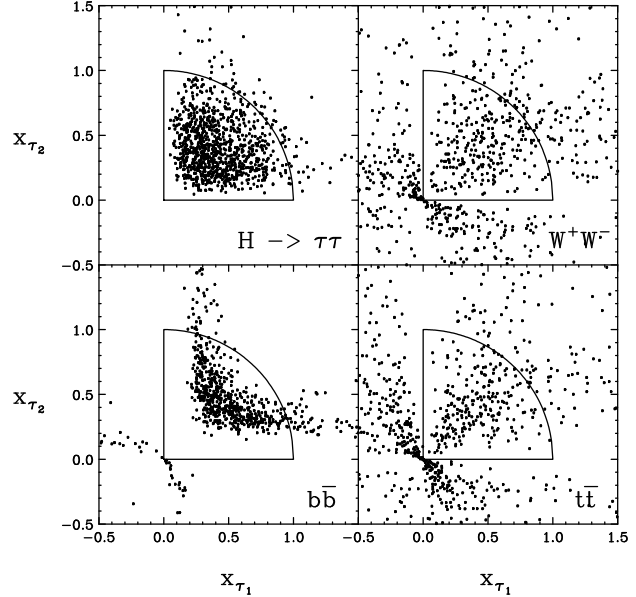} \vspace*{0.2cm}
\caption{Scatter plots of $x_{\tau_1}$ v. $x_{\tau_2}$ with the cuts of
Eqs.(\ref{eq:basic})-(\ref{eq:cosphi}), for the 120 GeV $Hjj$ signal, 
$b\bar{b}jj$, $WWjj$ and $t\bar{t} + jets$ reducible backgrounds.
The number of points in each plot is arbitrary and corresponds to
significantly higher integrated luminosities than expected for the LHC.
The solid lines indicate the cuts of Eq.~(\ref{eq:xtau}).}
\label{fig:x1x2}
\end{figure}

For significant further reduction of the various backgrounds, 
reconstruction of the tau-pair invariant mass~\cite{tautaumass} is necessary. 
Due to the large mass of the decaying Higgs boson and also because of 
its large transverse momentum (see Fig.~\ref{fig:ptH}) the produced taus 
are moving relativistically in the laboratory frame. As a result the tau 
direction closely follows the direction of the corresponding observed
decay lepton. Since the transverse momentum of the Higgs boson is known
(it is given by the vectorial sum of charged lepton $p_T$'s and missing
transverse momentum) the momentum parallelogram in the transverse plane
allows one to extract the fractions of the two tau momenta which are carried 
by the two charged leptons. We denote these momentum fractions by 
$x_{\tau_1},x_{\tau_2}$ in the following. This reconstruction works only 
if the taus are not emitted back-to-back in the transverse plane and we 
therefore impose the technical cut 
\begin{equation}
\label{eq:cosphi}
\cos{\phi_{e\mu}} \; > \; -0.9 \; .
\end{equation}
The resulting $x_{\tau_1},x_{\tau_2}$ distributions are shown in the 
form of a scatter plot of unweighted events in Fig.~\ref{fig:x1x2}. The 
distribution for real tau pairs is shown only for the Higgs boson signal
because the plot for $\gamma^*/Z\to\tau\tau$ looks virtually 
identical.

For real $\tau$ decays, the $\sla{p_T}$ vector must lie between the two 
leptons, and apart from finite detector resolution the reconstruction must 
yield $0 < x_{\tau_{1,2}} < 1$. For the $WW$ and $t\bar t$ backgrounds, 
however,
the collinear approximation is not valid because the $W$'s and top quarks 
receive only modest boosts in the lab. In this case, the $\sla{p_T}$ vector 
will rarely lie between the two leptons, and an attempt to reconstruct a 
$\tau$ pair will result in $x_{\tau_1} < 0$ or $x_{\tau_2} < 0$ for a 
significant fraction of the events~\cite{RZ_WW}. Many others end up in the
unphysical region $x_\tau>1$. The scatter plot of Fig.~\ref{fig:x1x2}
suggests the real $\tau$-reconstruction cuts
\begin{equation}
\label{eq:xtau}
x_{\tau_1} , \; x_{\tau_2} > 0 \; , \; \; \;
x_{\tau_1}^2 + x_{\tau_2}^2 < 1 \; .
\end{equation}

\begin{figure*}[hbt]
\begin{center}
\includegraphics[width=7.0cm,angle=90]{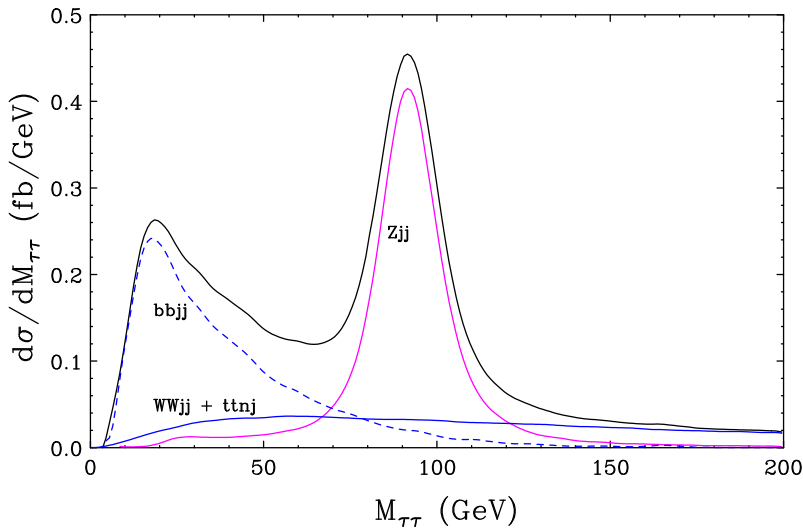}
\end{center}  \vspace*{-0.2cm}
\caption{Reconstructed tau-pair mass distribution for $WWjj$ and 
$t\bar{t}+jets$ events (solid blue), $b\bar bjj$ events (dashed blue) 
and QCD+EW $Zjj$ events (magenta). The combined curves are 
also shown (black).
The cuts of Eqs.~(\ref{eq:basic})-(\ref{eq:cosphi}) are imposed.}
\label{fig:earlyMtautau}
\end{figure*}

Once the momentum fractions carried by the $e,\mu$-pair are known, the 
invariant mass of the tau-pair is given by 
\begin{equation}
M_{\tau\tau} = m_{e\mu}/\sqrt{x_{\tau_1}x_{\tau_2}} \; .
\end{equation}
The reconstructed tau-pair invariant mass distributions for the combined 
$W^+W^-$ and $t\bar t$ backgrounds, for $b\bar bjj$ events and for the QCD 
and EW $Zjj$ events are shown in Fig.~\ref{fig:earlyMtautau}, after the 
back-to-back cut of Eq.~(\ref{eq:cosphi}). The $b\bar bjj$ background is 
almost completely concentrated in the small $M_{\tau\tau}$ region and 
$\gamma^*/Z+jj$ events are strongly peaked at $M_{\tau\tau} = M_Z$. When 
searching for a Higgs boson mass peak well above $M_Z$, both backgrounds 
are drastically reduced. As is indicated by the width of the $Z$ peak in 
Fig.~\ref{fig:earlyMtautau}, a resolution of about $10\%$ is possible for 
the reconstructed tau-pair invariant mass, which agrees well with earlier 
results obtained with full detector simulations for $A\to\tau\tau$ by 
ATLAS~\cite{Cavalli}. Here we are interested in SM Higgs bosons with a 
mass in the range 100~GeV~$< M_H <$~150~GeV. As a result we need to 
consider only backgrounds which lead to a reconstructed $M_{\tau\tau}$ in 
the range 
\begin{equation}
\label{eq:taumass}
90\ {\rm GeV}\; < M_{\tau\tau} < \; 160\ {\rm GeV} \, .
\end{equation}
The reduced background level due to this tau-pair mass cut is shown 
in the last line of Table~\ref{taudata_all}. 

\begin{figure}[tb]
\includegraphics[width=11.8cm,angle=90]{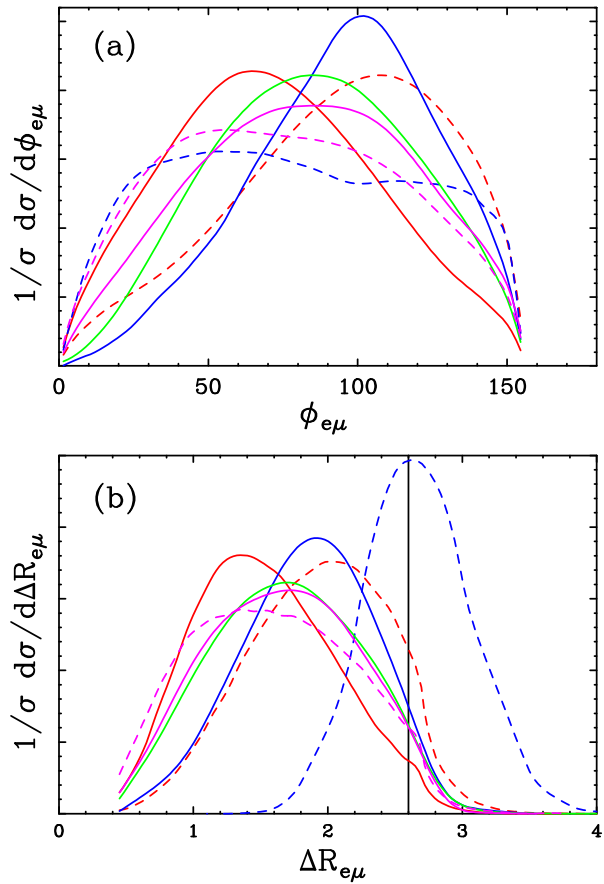} \vspace*{0.2cm} 
\caption{Normalized angular distributions of the charged leptons: 
a) azimuthal opening angle and b) separation in the lego plot. 
Results are shown for Higgs boson masses of 100 and 150~GeV (solid and 
dashed red lines, respectively) and for the various backgrounds: 
$t\bar{t} + jets$ (solid blue), $b\bar{b}jj$ (dashed blue), 
combined QCD and EW $WWjj$ (solid green), 
QCD $\tau\tau jj$ (solid magenta) and 
EW $\tau\tau jj$ (dashed magenta). 
The cuts of Eqs.~(\ref{eq:basic})-(\ref{eq:xtau}),(\ref{eq:taumass}) 
are imposed.}
\label{fig:angdist}
\end{figure}

Clearly, Higgs mass reconstruction is a very powerful background suppression 
tool, in particular for $b\bar bjj$ events which mostly populate the low 
$M_{\tau\tau}$ region. This means that the background cross sections in 
Table~\ref{taudata_all} exaggerate the background level and one should rather 
consider the expected rates in the vicinity of the Higgs boson mass peak.
Given the expected mass resolution, we only need to consider background 
events within $\pm 10$~GeV of the peak. In Table~\ref{taudata} we have 
summarized these cross sections at the various cut levels for the example of 
a Higgs boson at $M_H = 120$~GeV. 
Within the cuts of Eqs.~(\ref{eq:basic})-(\ref{eq:xtau})
we have achieved a signal to background (S/B) ratio of 1/1. 

Yet another significant difference between signal and some 
backgrounds is the angular distribution of the charged decay 
leptons, $e^\pm$ and $\mu^\mp$, relative to each other. In the case of the 
Higgs signal, the high $p_T$ of the Higgs boson results in a tau pair, and
therefore charged decay leptons, which are 
emitted fairly close together in the lab frame. In the case of the 
heavy quark backgrounds, this correlation is not reproduced, in particular
when viewed as lepton separation in the lego plot 
(see Fig.~\ref{fig:angdist}). For the $b\bar bjj$ background, for example, 
a large rapidity separation is induced by the conflicting requirements of a
large tau-pair invariant mass and the low lepton transverse 
momenta surviving the lepton isolation cuts. The lepton correlations can 
be exploited by imposing a lepton pair angular cut:
\begin{equation}
\label{eq:ang}
\triangle R_{e\mu} < 2.6 \, .
\end{equation}
This cut acts primarily against the $t\bar{t}+jets$ and $b\bar{b}jj$ 
backgrounds, which are already at a quite low level. We select the value 
2.6 conservatively to retain more signal rate, in particular for large 
Higgs boson masses, close to $M_H = 150$~GeV.

At this level of cuts we consider one final background, an additional source 
of $e + \mu + \sla{p}_T$ from Higgs production itself, via $H\to WW$ decay. 
Real or slightly virtual $W$'s are produced as opposed to real $\tau$'s, 
so the search for real 
$\tau$'s outlined above will restrict the contribution from this decay 
channel. However, most of the other cuts we have described isolate Higgs 
production only, and even the lepton angular cut will select $H\to WW$ 
events due to the strong anti-correlation of the $W$ spins, which leads to 
the $e,\mu$ pair being emitted preferentially together in the rest frame 
of the Higgs boson~\cite{DittDrein}; the large transverse boost of the Higgs 
boson in the lab only enhances this angular correlation~\cite{RZ_WW}. 
The large $WW$ branching ratio compared to that for $\tau\tau$ 
over the upper end of the mass range which we are considering 
($\approx 130-150$~GeV), then leads to a background component which 
cannot be neglected.
Table~\ref{HvH} demonstrates this via a comparison of the $H\to\tau\tau$ 
signal and $H\to WW$ background over the mass range of interest, after all 
cuts previously discussed have been imposed.

\begin{figure}[tb]
\includegraphics[width=6.5cm,angle=90]{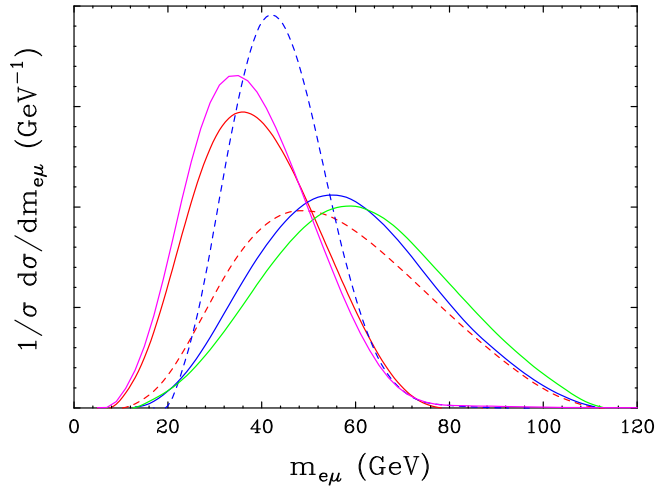} \vspace*{0.2cm} 
\caption{Normalized dilepton invariant mass distribution, after the cuts of 
Eqs.~(\ref{eq:basic})-(\ref{eq:xtau}), (\ref{eq:taumass}),(\ref{eq:ang}) 
have been imposed. 
Results are shown for Higgs boson masses of 100 and 150~GeV (solid and dashed 
red lines, respectively) and 
for the various backgrounds as in Fig.~\ref{fig:ptmiss}, except that the QCD 
and EW $\tau\tau jj$ backgrounds have been combined for clarity 
(solid magenta line), and the solid green line corresponds to the combined 
QCD + EW $WWjj$ background.}
\label{fig:mll}
\end{figure}

Another distribution of interest is the lepton-pair invariant mass, 
$m_{e\mu}$, which is shown in Fig.~\ref{fig:mll} for Higgs masses of 
$M_H = 100$~GeV and 150~GeV and for the various backgrounds. As is readily 
seen, the $t\bar{t} + jets$ and $WWjj$ backgrounds prefer high values of
$m_{e\mu}$, while the real-tau backgrounds cluster at low invariant 
mass.  A cut on this observable would have to be Higgs mass specific. 
We therefore do not cut on $m_{e\mu}$, but instead mention it as an additional 
distinguishing characteristic of the Higgs signal at a very advanced level 
of cuts.

\begin{table*}
\caption{
Signal rates $\sigma\cdot B(H\to\tau\tau\to e^\pm\mu^\mp\sla{p_T})$ for 
$M_H = 120$~GeV 
and corresponding background cross sections, in $pp$ collisions at 
$\protect\sqrt{s}=14$~TeV. Results are given for various levels of cuts and 
are labeled by equation numbers discussed in the text. No constraint on 
the reconstructed $\tau\tau$ invariant mass is imposed, except in
the last line which requires 90~GeV$<m_{\tau\tau}<$160~GeV. 
All rates are given in fb.}
\vspace{0.15in}
\label{taudata_all}
\begin{tabular}{l|ccccccc|c}
     &       & QCD           & EW            &
     &              & QCD    & EW     &     \\
cuts & $Hjj$ & $\tau\tau jj$ & $\tau\tau jj$ & $t\bar{t} + jets$ 
     & $b\bar{b}jj$ & $WWjj$ & $WWjj$ & S/B \\
\hline
forward tags (\ref{eq:basic})-(\ref{eq:gap})
& 2.2  & 57   & 2.3  & 1230 & 1050 & 4.9   & 3.3   & 1/1100 \\
+ $b$ veto (\ref{eq:bveto})
&      &      &      & 72   &      &       &       & 1/550  \\
+ $\sla{p}_T$ (\ref{eq:ptmiss})
& 1.73 & 29   & 1.57 & 62   & 29   & 4.1   & 2.9   & 1/74   \\
+ $M_{jj}$ (\ref{eq:mjj})
& 1.34 & 10.3 & 1.35 & 16.3 & 10.4 & 1.60  & 2.6   & 1/32   \\
+ non-$\tau$ reject. (\ref{eq:cosphi},\ref{eq:xtau})
& 1.15 & 5.2  & 0.63 & 0.31 & 0.42 & 0.032 & 0.042 & 1/5.8  \\
\end{tabular}
\end{table*}

\begin{table*}
\caption{
Signal rates $\sigma\cdot B(H\to\tau\tau\to e^\pm\mu^\mp\sla{p_T})$ for a 
SM Higgs boson of $M_H = 120$~GeV and corresponding background cross sections, 
within $\pm 10$~GeV mass bins. Results are given for 
various levels of cuts and are labeled by equation numbers discussed in the 
text. On line seven we include an overall efficiency factor for 
identification of tagging jets and leptons, as discussed in the text. On 
line eight the minijet veto is included, with $p_T^{veto} = 20$~GeV. 
All rates are given in fb.}
\vspace{0.15in}
\label{taudata}
\begin{tabular}{l|cccccccc|c}
     & $H\to\tau\tau$ & $H\to WW$ & QCD & EW & & & QCD & EW & \\
cuts & signal & bkgd & $\tau\tau jj$ & $\tau\tau jj$ 
     & $t\bar{t} + jets$ & $b\bar{b}jj$ & $WWjj$ & $WWjj$ & S/B \\
\hline
forward tags (\ref{eq:basic}-\ref{eq:gap})
& 1.34 &       & 4.7  & 0.18 & 45  & 8.2  & 0.18  & 0.11  &  1/44 \\
+ $b$ veto (\ref{eq:bveto})
&      &       &      &      & 2.6 &      &       &       &  1/12 \\
+ $\sla{p}_T$ (\ref{eq:ptmiss})
& 1.17 &       & 2.3  & 0.12 & 2.0 & 0.28  & 0.12  & 0.08  & 1/4.1 \\
+ $M_{jj}$ (\ref{eq:mjj})
& 0.92 &       & 0.67  & 0.10  & 0.53  & 0.13  & 0.049 & 0.073 & 1/1.7 \\
+ non-$\tau$ reject. (\ref{eq:cosphi},\ref{eq:xtau})
& 0.87 &       & 0.58  & 0.10  & 0.09  & 0.10  & 0.009 & 0.012 & 1/1   \\
+ $\triangle R_{e\mu}$ (\ref{eq:ang})
& 0.84 & 0.023 & 0.52  & 0.086 & 0.087 & 0.028 & 0.009 & 0.011 & 1.1/1 \\
+ ID effic. (${\it\times 0.67}$)
& 0.56 & 0.015 & 0.34  & 0.058 & 0.058 & 0.019 & 0.006 & 0.008 & 1.1/1 \\
$P_{surv,20}$ & ${\it\times 0.89}$ & ${\it\times 0.89}$ 
              & ${\it\times 0.29}$ & ${\it\times 0.75}$ 
              & ${\it\times 0.29}$ & ${\it\times 0.29}$ 
              & ${\it\times 0.29}$ & ${\it\times 0.75}$ & - \\
+ minijet veto (\ref{eq:mjveto})
& 0.50 & 0.014 & 0.100 & 0.043 & 0.017 & 0.006 & 0.002 & 0.006 & 2.7/1 \\
\end{tabular}
\end{table*}

\begin{table*}
\caption{Cross section times branching ratio for the $H\to\tau\tau$ signal 
vs. $H\to WW$ background, for the mass range of interest. The cuts of 
Eqs.~(\ref{eq:basic})-(\ref{eq:xtau}),(\ref{eq:ang}) are imposed, and rates
correspond to $m_{\tau\tau}=M_H\pm 10$~GeV mass bins around the Higgs 
boson mass.}
\vspace{0.15in}
\label{HvH}
\begin{tabular}{l|ccccccccccc}
$M_H$ & 100   & 105   & 110  & 115   & 120   & 125   & 130   & 135   & 140   & 145   & 150   \\
\hline
B($H\to\tau\tau$)$\cdot\sigma$ (fb)
      & 1.04  & 1.03  & 0.98 & 0.93  & 0.84  & 0.74  & 0.62  & 0.51  & 0.39  & 0.27  & 0.19  \\
B($H\to WW$)$\cdot\sigma$ (fb)
      & 0.002 & 0.005 & 0.00 & 0.015 & 0.024 & 0.034 & 0.045 & 0.057 & 0.067 & 0.072 & 0.076 \\
\end{tabular}
\end{table*}

While we do not impose any further cuts at this point, we should include an 
estimate of the total rate loss due to various detector efficiencies, to 
make closer contact with experimental expectations. Based on discussions 
with ATLAS and CMS experimentalists, we apply an additional factor $0.86^2$ 
for the ID efficiency of the two tagging jets, and a factor $0.95^2$ for 
the ID efficiency of the two charged leptons, $e$ and $\mu$. The combined
detection efficiencies are
reflected in line 7 of Table~\ref{taudata}. We note that the high 
efficiency for lepton triggering and identification may not hold for 
all leptons down to $p_T = 10$~GeV, but we do not expect losses to be 
large enough to render our estimate grossly optimistic.

\begin{figure}[tb]
\includegraphics[width=12.0cm,angle=90]{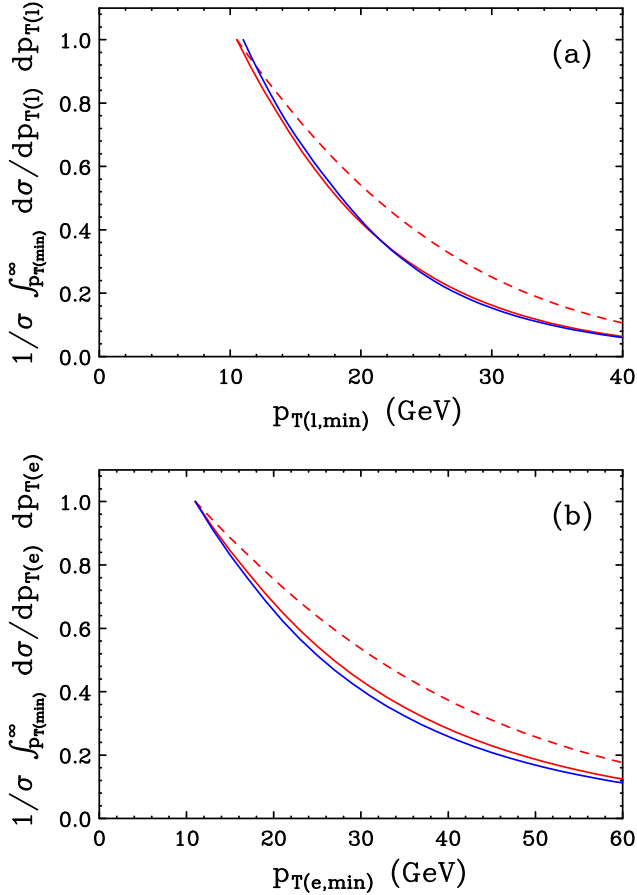} \vspace*{0.2cm}
\caption{Integrated charged lepton $p_T$ distributions after the cuts of
Eqs.~(\ref{eq:basic})-(\ref{eq:xtau}),(\ref{eq:taumass}),(\ref{eq:ang})
have been imposed. Shown are events fractions above a minimal lepton $p_T$ as
compared to a threshold of 10~GeV for (a) the minimal $e$ or $\mu$ transverse 
momentum, $p_{T_{\ell,min}}$, and (b) the electron transverse momentum, 
$p_{T,e}$. Results are shown for Higgs boson masses of $M_H = 100$~GeV
(solid red line) and 150~GeV (dashed red line) and for the combined 
background (blue line).
}
\label{fig:ptl}
\end{figure}

The consequences of higher effective lepton transverse momentum thresholds 
are explored in Fig.~\ref{fig:ptl}. An increase of both the electron and the 
muon threshold to e.g. $p_{T,\ell}> 15$~GeV would lead to a signal loss of
order $30\%$, with a slight improvement of S/B, as can be read off 
Fig.~\ref{fig:ptl}(a). An increase of only the electron
threshold to this value would reduce the $H\to\tau\tau$ signal by less
than $20\%$ (see Fig.~\ref{fig:ptl}(b)). $p_T$ thresholds may be as 
high as 20~GeV for electrons but can be as low as 6~GeV for triggering on
muons~\cite{CMS-ATLAS,cernsims}. Given the complex nature of the signal
at hand, containing electrons, muons, missing transverse momentum and jets,
the issue of devising an effective trigger, at low and high luminosity, 
needs to be addressed within  
a full detector study and cannot be performed by us. We want to emphasize 
the premium for low lepton $p_T$ thresholds, however.


\section{Radiation patterns of minijets}
\label{sec:minijet}

If we are to veto central $b$ jets to reduce the $t\bar{t} + jets$ background 
to a manageable level, we must take care to correctly estimate higher-order 
additional central partonic emission in the signal and 
backgrounds~\cite{RZ_WW}. 
Fortunately, due to the absence of color exchange between the two scattering 
quarks in EW processes, which includes our $Hjj$ signal, we expect soft gluon 
emission mainly in the very forward and very backward directions. For 
QCD processes, on the other hand, which are dominated by $t$-channel color 
octet exchange, soft gluon radiation occurs mainly in the central detector. 
Thus, when considering  
additional central radiation with $p_T \geq 20$~GeV to match our $b$ veto 
condition, we will reject QCD backgrounds with much higher probability 
than the EW processes. Our $b$ veto is then automatically also a minijet veto, 
a tool for QCD background suppression which has been previously studied in 
great detail for $Hjj$ production at hadron 
colliders~\cite{bpz_mj,thesis,RZ_tautau,DZ_IZ_minijet}. 

Largely following the analysis of Ref.~\cite{RSZ_vnj} for the 
analogous EW $Zjj$ process which would be used to ``calibrate'' the tool at 
the LHC, we veto additional central jets in the region
\begin{mathletters}
\label{eq:mjveto}
\ba
p_{Tj}^{\rm veto} \; & > & \; p_{T,{\rm veto}}\;, \label{eq:ptveto} \\
\eta_{j,min}^{\rm tag} \; & < & \eta_j^{\rm veto}
< \; \eta_{j,max}^{\rm tag} \; , \label{eq:etaveto}
\ea
\end{mathletters}
where $p_{T,\rm veto}$ may be chosen based on detector capabilities and
expected minijet production from double parton scattering. Here we take
$p_{T,{\rm veto}}=20$~GeV.

The determination of veto efficiencies for the QCD backgrounds 
encounters the problem that we are interested in the phase space region 
where additional soft parton emission is very likely. In a fixed order 
perturbative calculation of cross sections with an additional soft
central jet of $p_{Tj}>p_{T,\rm veto}$ 
this leads to a ``3-jet'' cross section (counting the two 
forward tagging jets plus the soft central veto-jet) which often exceeds
the corresponding hard ``2-jet'' cross section considered in the previous
Section. The occurrence of $\sigma_{3\;jet}>\sigma_{2\;jet}$ indicates
that effects of multiple soft gluon emission must be taken into account.
We have analyzed this problem in detail 
before~\cite{RZ_tautau,RSZ_vnj,RZ_WW,thesis} and found that two very
different procedures give consistent results for the veto probabilities.
The first, the truncated shower approximation (TSA)~\cite{TSA}, 
regularizes the soft parton $p_T$ distribution so as to reproduce the 
$p_T$ distribution of the hard recoil system which is expected from a full 
resummation calculation, while preserving the normalization of the hard
2-jet cross section. The second, the ``exponentiation model'' assumes 
that soft gluon radiation approximately exponentiates. This implies that 
central veto jet multiplicities effectively follow a Poisson distribution
with mean $\bar n_{jet} = \sigma_3/\sigma_2$~\cite{CDFjets}. 
Both methods use perturbative 
cross sections for 2- and 3-jet cross sections. The advantage is that 
QCD matrix elements at tree level contain the full information on angular
distributions and hardness of additional jet emission. A parton shower 
approach would not immediately give reliable answers unless both color 
coherence and the choice of scale are implemented correctly, matching the 
answer given by QCD matrix elements for sufficiently hard partons.

In the following we directly use the results of Ref.~\cite{RZ_WW,thesis}, 
which considered Higgs production by WBF and background processes in 
phase space regions for the jets that are virtually identical to the 
ones considered here. In contrast to our early 
studies~\cite{RZ_tautau,RSZ_vnj}, the veto 
candidates are defined jets ($p_T > 20$~GeV) anywhere between the tagging 
jets, i.e. they are searched for in a somewhat larger rapidity region than 
the $\tau$ decay leptons (see Eq.~(\ref{eq:lepcen})), which have to be at 
least 0.7 units of rapidity away from the tagging jets. 
The choice of Eq.~(\ref{eq:etaveto}) allows for more suppression of the 
backgrounds than the more restrictive selection. 

The resulting veto 
survival probabilities are summarized in line eight of Table~\ref{taudata}.
These values were determined with 3-jet Monte Carlo simulations for the 
Higgs signal and QCD and EW $Zjj$ production~\cite{RZ_WW,thesis}. 
The results for QCD $Zjj$ production are also used for QCD $WWjj$ production, 
due to the similarity of the subprocesses, and are taken as well for the 
$b\bar{b}jj$ background. 

For the $t\bar{t} + jets$ backgrounds we have discussed before, the effects 
of the minijet veto on the $b$-quarks arising from the top quark decays has 
been considered (see e.g. the second line of Table~\ref{taudata}). We now 
want to take into account the additional reduction due to soft gluon 
radiation in $t\bar t$ events. Note that this separation is an artifact of 
our using tree level approximations and would not arise in either a NLO 
calculation or when using a parton shower program (and certainly not in the 
experiments). We have examined the expected survival probability for  
$t\bar{t} + jets$ for the exponentiation model in Refs.~\cite{RZ_WW,thesis}, 
finding a somewhat higher veto probability than for $Zjj$ events.
Because of the uncertainties in the determination of veto probabilities 
we prefer the more conservative estimates here. Note, however, that even 
without considering any additional minijets, the $t\bar t$ backgrounds are 
very small (see line seven of Table~\ref{taudata}), mitigating any concern 
over minijet veto probabilities. 
While the veto survival probabilities of Table~\ref{taudata} are estimates 
only, they can be independently determined at the LHC in processes like 
$Zjj$ and $Wjj$ production~\cite{RSZ_vnj,CZ_gap}. 

\begin{figure*}[htb]
\begin{center}
\includegraphics[width=16cm]{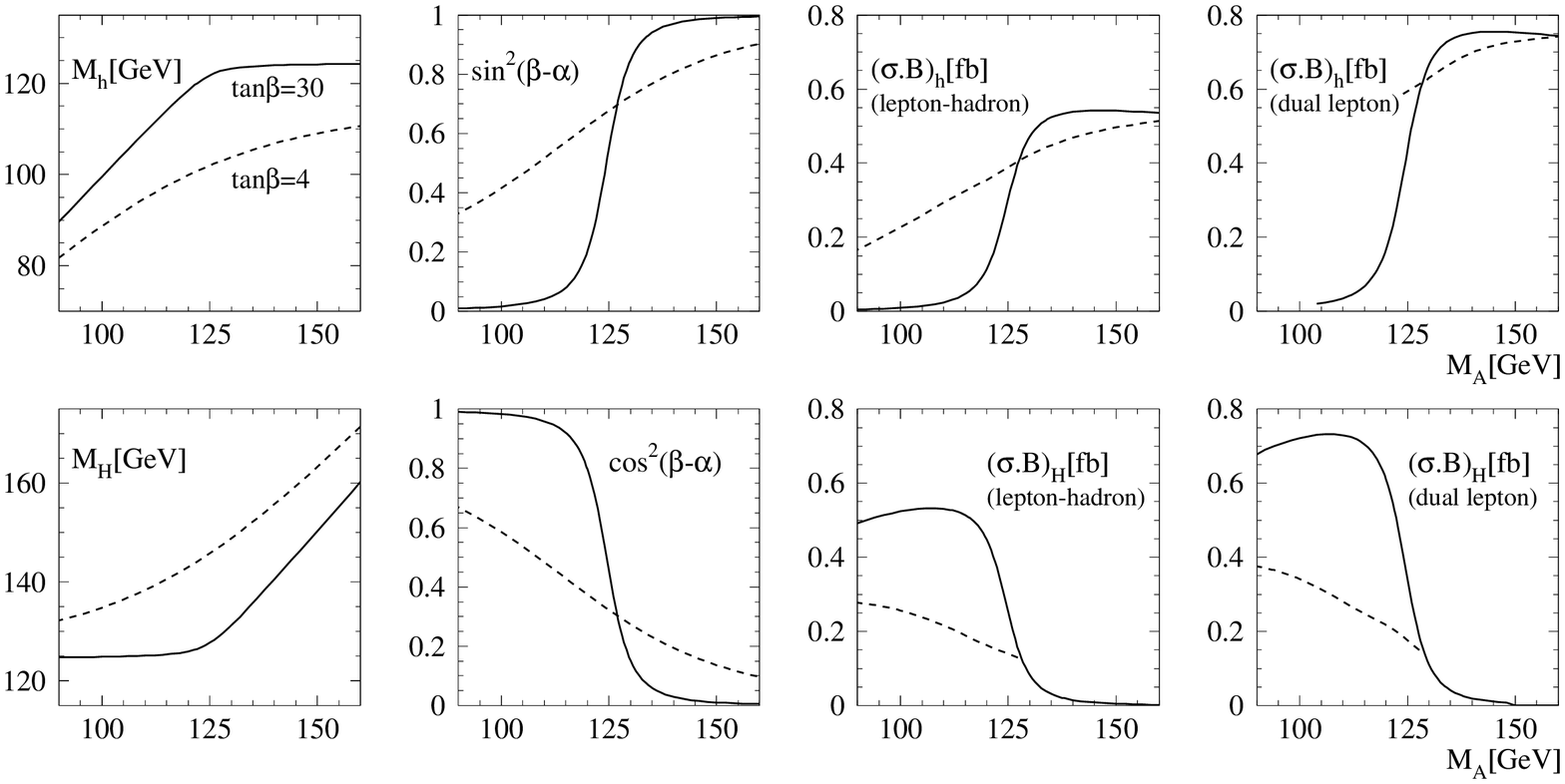} 
\end{center} \vspace*{-0.2cm}
\caption[]{Variation of Higgs boson masses, couplings to gauge bosons,
  and signal rates for the two $\tau\tau$ channels as a function of
  the pseudoscalar Higgs boson mass. The complementarity of the two
  scalar plateau states is shown for $\tan\beta=4,30$. Other MSSM
  parameters are set to $\mu=200$~GeV, $M_{\rm SUSY}=1$~TeV, and
  maximal mixing.}
\label{fig:susycoup}
\end{figure*}

So far we have considered a single Higgs boson mass of 120~GeV only. 
We must extend our results to a larger range of $M_H$. 
The expected number of signal and background events for 
100~GeV~$\leq M_H \leq 150$~GeV and an integrated luminosity of 60~fb$^{-1}$ 
are shown in Table~\ref{summary}. In the fourth line of Table~\ref{summary} 
we show the S/B rate, and in the fifth line we show the Poisson 
probability for the combined backgrounds to fluctuate up to the signal 
level, in terms of the equivalent Gaussian significances which can be 
expected in the experiment on average. Low luminosity running is assumed, 
i.e. no efficiency losses due to overlapping minimum bias events are 
considered.


\section{Implications for the MSSM}
\label{sec:MSSM}

The $H\to\tau\tau$ decay mode has proven especially useful in case of the 
minimal supersymmetric extension of the Standard Model~\cite{PRZ_mssm}. 
In the MSSM the neutral CP even Higgs boson states form two mass eigenstates, 
the properties of which are usually described as functions of $\tan\beta$ 
and the pseudoscalar mass $M_A$. For very large or very small values of 
$M_A$ the scalar masses $M_h$ or $M_H$ approach a plateau, as shown in 
Fig.~\ref{fig:susycoup}. These plateau states have masses below 
$\lesssim 130$~GeV~\cite{one_loop,two_loop}, dependent on the value of 
$\tan\beta$. Together with the projected limits from the $Zh$ and $ZH$ 
search at LEP2 this yields exactly the mass window where the WBF 
$H\to\tau\tau$ mode is most promising, as shown in Table~\ref{summary}.

\begin{figure}[tb]
\includegraphics[width=8.9cm]{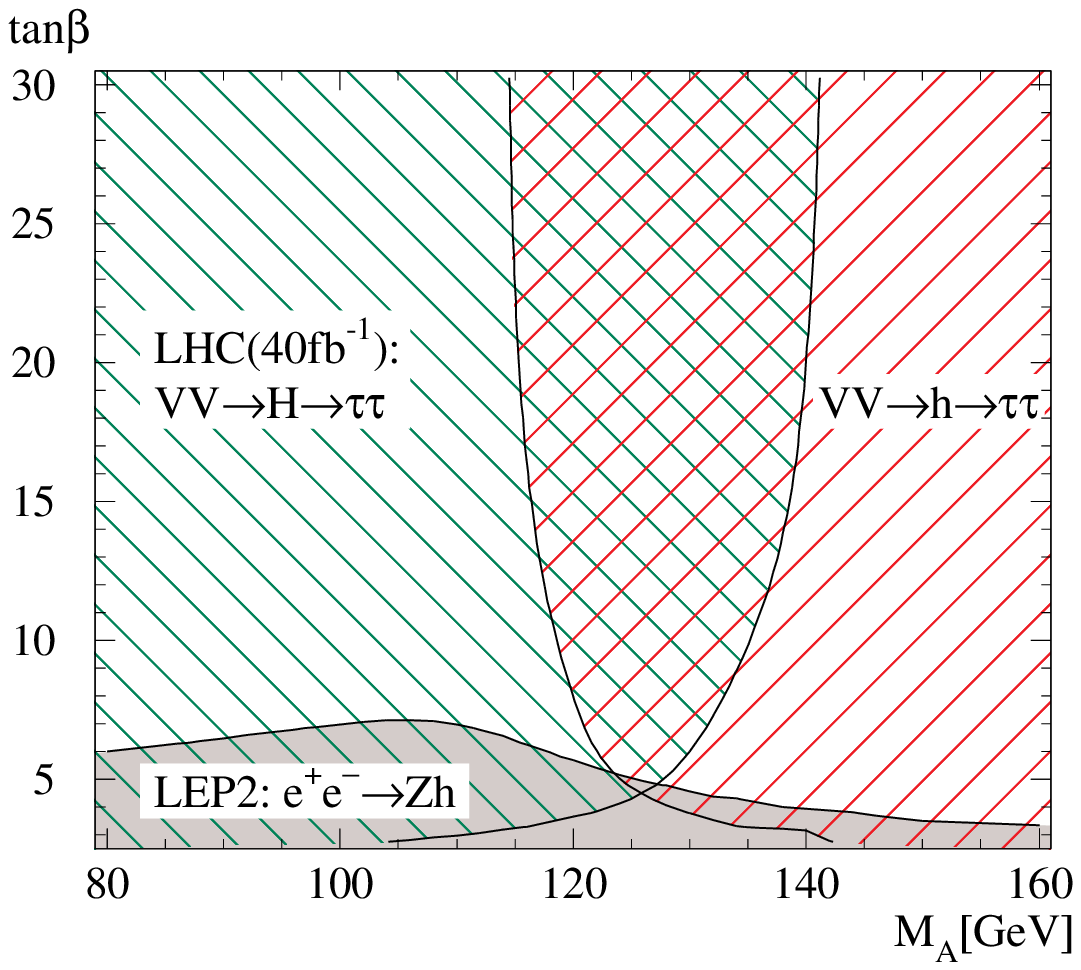} 
\includegraphics[width=8.9cm]{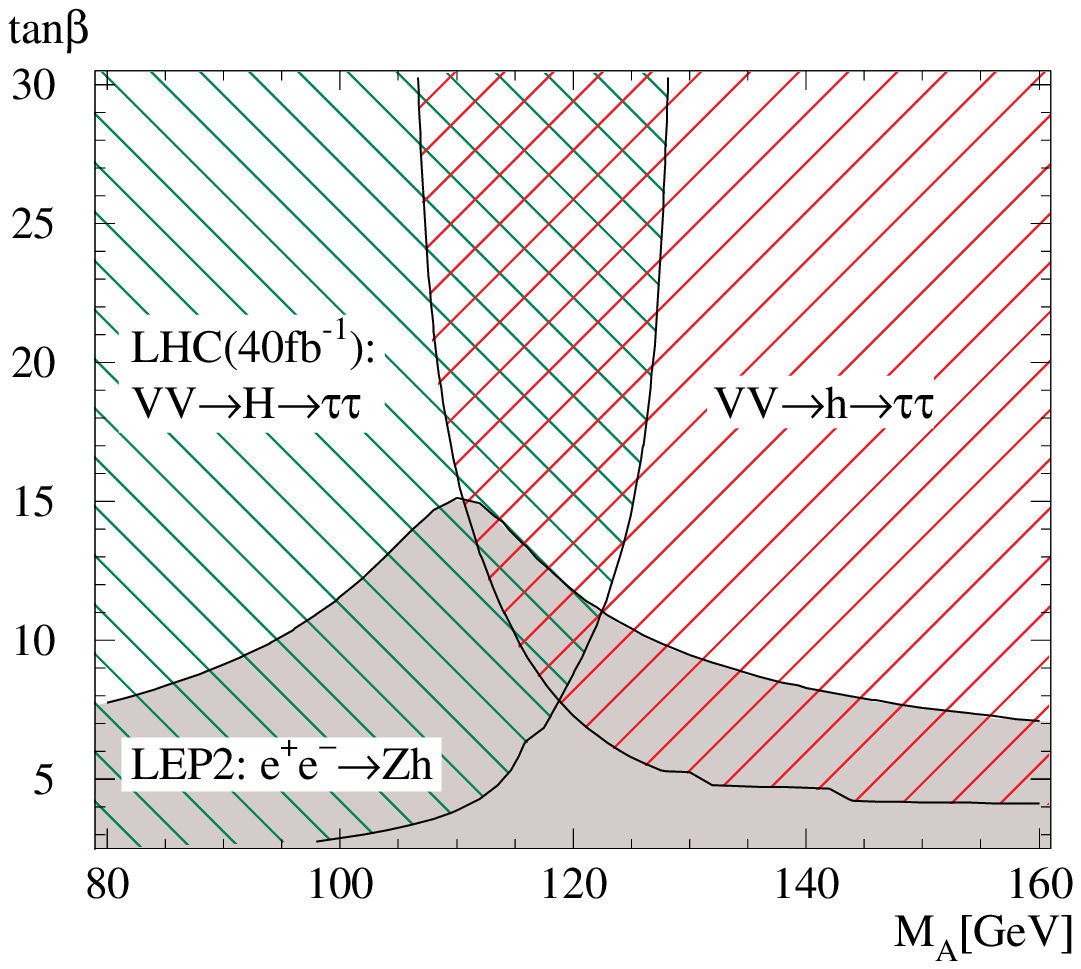} \vspace*{0.2cm}
\caption[]{$5 \sigma$ discovery contours for $h,H\to\tau\tau$ at the 
  LHC with $\lesssim~$40~fb$^{-1}$. They are complemented by the
  current LEP2 limits. The SUSY parameters are set to
  $\mu=200$~GeV, $M_{\rm SUSY}=1$~TeV, and maximal mixing (upper) and
  no mixing (lower).}
\label{fig:susyreach}
\end{figure}

The couplings of the MSSM Higgs bosons to $W,Z$ and $\tau$ pairs are given
by the SM coupling and a multiplicative SUSY factor which for 
the weak bosons are
\begin{align}
\eta_{VVh} &= \sin(\beta-\alpha) \, , \notag \\
\eta_{VVH} &= \cos(\beta-\alpha) \, , 
\end{align}
while for down-type fermions the factors are
\begin{align}
\eta_{\tau\tau h} &= -\frac{\sin\alpha}{\cos\beta} \notag \\
                  &= \, \sin(\beta-\alpha) - \tan\beta \; \cos(\beta-\alpha)
\, , \notag \\
\eta_{\tau\tau H} &= \, \frac{\cos\alpha}{\cos\beta} \notag \\
         &= \, \cos(\beta-\alpha) + \tan\beta \; \sin(\beta-\alpha) \, ,
\label{eq:susycoup}
\end{align}
where $\alpha$ is the mixing angle between the two CP even Higgs boson 
states. On the $M_h$ and $M_H$ plateaus one finds
$\alpha \approx \beta-\pi/2$ and $\alpha \approx -\beta$, leading to 
$\eta_{VVh} \approx 1$ on the $M_h$ plateau and $\eta_{VVH} \approx 1$ on 
the $M_H$ plateau (see Fig.~\ref{fig:susycoup}). The $\tau\tau$ SUSY 
factors on the plateaus are dominated by the first terms in 
Eq.~(\ref{eq:susycoup}), again rendering unity.\footnote{The case 
of vanishing $\eta_{\tau\tau h}$ or $\eta_{\tau\tau H}$ has previously 
been discussed in detail~\cite{PRZ_mssm}: $h,H \to\gamma\gamma$ would 
be dramatically enhanced and become the dominant discovery mode.} 
Thus, in both plateau regions the situation is very similar to the
SM case discussed above.
As illustrated in Fig.~\ref{fig:susycoup} these plateau states are 
numerically approached for moderate $M_A$ values already, as long as 
$\tan\beta$ does not become too small, 
which would be in conflict with LEP2 limits. In fact, the 
MSSM branching ratio $B(h,H \to\tau\tau)$ is even slightly enhanced 
compared to the SM case~\cite{hdecay}.

Previously we have analyzed the semileptonic $\tau\tau$ decay mode
$h/H\to \tau\tau \to e/\mu^\pm h^\mp\sla p_T$~\cite{PRZ_mssm}. To the
semileptonic channel we can now add the signal expected in the 
purely leptonic mode, $h/H\to\tau\tau\to e^\pm\mu^\mp\sla{p_T}$.
In this extended analysis we change our definition of the 
``MSSM Higgs boson signal'' as compared to the previous analysis, which 
yielded a $5\sigma$ coverage of the entire parameter space with 
$\approx 100$~fb$^{-1}$ luminosity. In the WBF $\tau\tau$ mode the 
SM and the MSSM analyses are exactly the same, namely scanning the 
invariant $\tau\tau$ mass distribution for a Higgs mass peak and finding 
the probability that the excess events could arise from a fluctuation of the 
background within a $\pm 10$~GeV mass window.
For values of $M_A$ in the transition region the reconstructed 
mass peaks from $h\to\tau\tau$ and $H\to\tau\tau$ decays will be close to 
each other. Hence, the tail of the invariant mass distribution resulting 
from the non-plateau state will add to the plateau state signal, reducing 
the required luminosity for a complete coverage to $\approx 80$~fb$^{-1}$ 
for the semileptonic channel alone.

For each of the two discovery modes $j$ (semileptonic and purely leptonic)
we calculate the probability $P_j$ 
for the background to fluctuate and produce all expected signal+background 
events. $P_1$ and $P_2$ are combined to 
$P_1 P_2 (1-\log(P_1P_2))$~\cite{higgswg}. This probability instead of the 
na\"{\i}ve product of the two single channel probabilities is then 
translated into the luminosity required for a $5\sigma$ 
discovery.\footnote{Some analyses prefer to chose the Bayesian instead of 
the Frequentist approach, which would lead to a slight decrease in our 
required luminosity.}
However, since we consider two channels of similar strength we do not 
observe significant influence of the statistical treatment on the final 
numbers.

\begin{table*}[htb]
\caption{Number of expected events for a SM $Hjj$ signal in the 
$H\to\tau\tau\to e^\pm\mu^\mp\sla{p_T}$ channel, for a range of Higgs boson
masses. Results are given for 60~${\rm fb}^{-1}$ 
of data at low luminosity running, and application of all efficiency factors 
and cuts, including a minijet veto. 
As a measure of the Poisson probability of the background to fluctuate up to 
the signal level, the last line gives $\sigma_{Gauss}$, the number of 
Gaussian equivalent standard deviations.}
\vspace{0.15in}
\label{summary}
\begin{tabular}{c|ccccccccccc}
$M_H$ & 100  & 105  & 110  & 115  & 120  & 125  & 130  & 135  & 140  & 145  & 150  \\
\hline
$\epsilon\cdot\sigma_{sig}$ (fb)
      & 0.62 & 0.61 & 0.58 & 0.55 & 0.50 & 0.44 & 0.37 & 0.30 & 0.23 & 0.16 & 0.11 \\
$N_S$ & 37.4 & 36.5 & 35.0 & 32.8 & 30.0 & 26.3 & 22.3 & 18.0 & 13.7 &  9.9 &  6.5 \\
$N_B$ & 67.7 & 45.4 & 27.4 & 16.8 & 11.2 &  8.4 &  7.1 &  6.4 &  6.1 &  5.9 &  5.7 \\
S/B   &  0.6 &  0.8 &  1.3 &  2.0 &  2.7 &  3.2 &  3.1 &  2.8 &  2.2 &  1.7 &  1.1 \\
$\sigma_{Gauss}$ & 4.1 & 4.8 & 5.6 & 6.4 & 6.8 & 6.7 & 6.1 & 5.3 & 4.3 & 3.2 & 2.2 \\
\end{tabular}
\end{table*}

To estimate the required luminosity for complete coverage of the MSSM 
parameter space with an expected $5\sigma$ signal we chose the 
supersymmetric mass scale $M_{\rm SUSY}$ as 1~TeV and vary the trilinear 
stop coupling $A_t$ between 0~(no mixing) and $\sqrt{6} \; M_{\rm SUSY}$ 
(maximal mixing). The latter yields the maximal plateau mass. However, 
from Table~\ref{summary} we conclude that our limit is very robust 
against effects which move the plateau masses. As a complementary 
measurement small values of $\tan\beta$ are excluded through the $Zh,ZH$ 
limits from LEP2. This assures that the Higgs mass peak is sufficiently 
separated from the $Z$ background peak. We are aware that other LEP2 
channels like $Ah,AH$ or the Tevatron RunII search will probe a fraction 
of the MSSM parameter space, but we do not need to rely on them for complete
coverage of the MSSM parameter space. The results are shown in 
Fig.~\ref{fig:susyreach}: combining 
the two tau decay channels leads to a required luminosity of 
$\lesssim~$40~fb$^{-1}$ for a $5\sigma$ observation, with a comfortable 
overlap in the no-mixing scenario. This number still depends on the LEP2 
reach, which we conservatively fix to the current limits~\cite{lep}.
The transition region of moderate 
pseudoscalar masses (which limits the reach) exhibits one additional 
feature: if the luminosity is large enough there will be a growing region 
where both the light and the heavy scalar will appear as peaks in the 
invariant mass spectrum. This would lead to a unique opportunity to 
measure the mixing angles in the MSSM Higgs sector.


\section{Discussion}
\label{sec:disc}

\begin{figure*}[tbh]
\begin{center}
\includegraphics[width=7.0cm,angle=90]{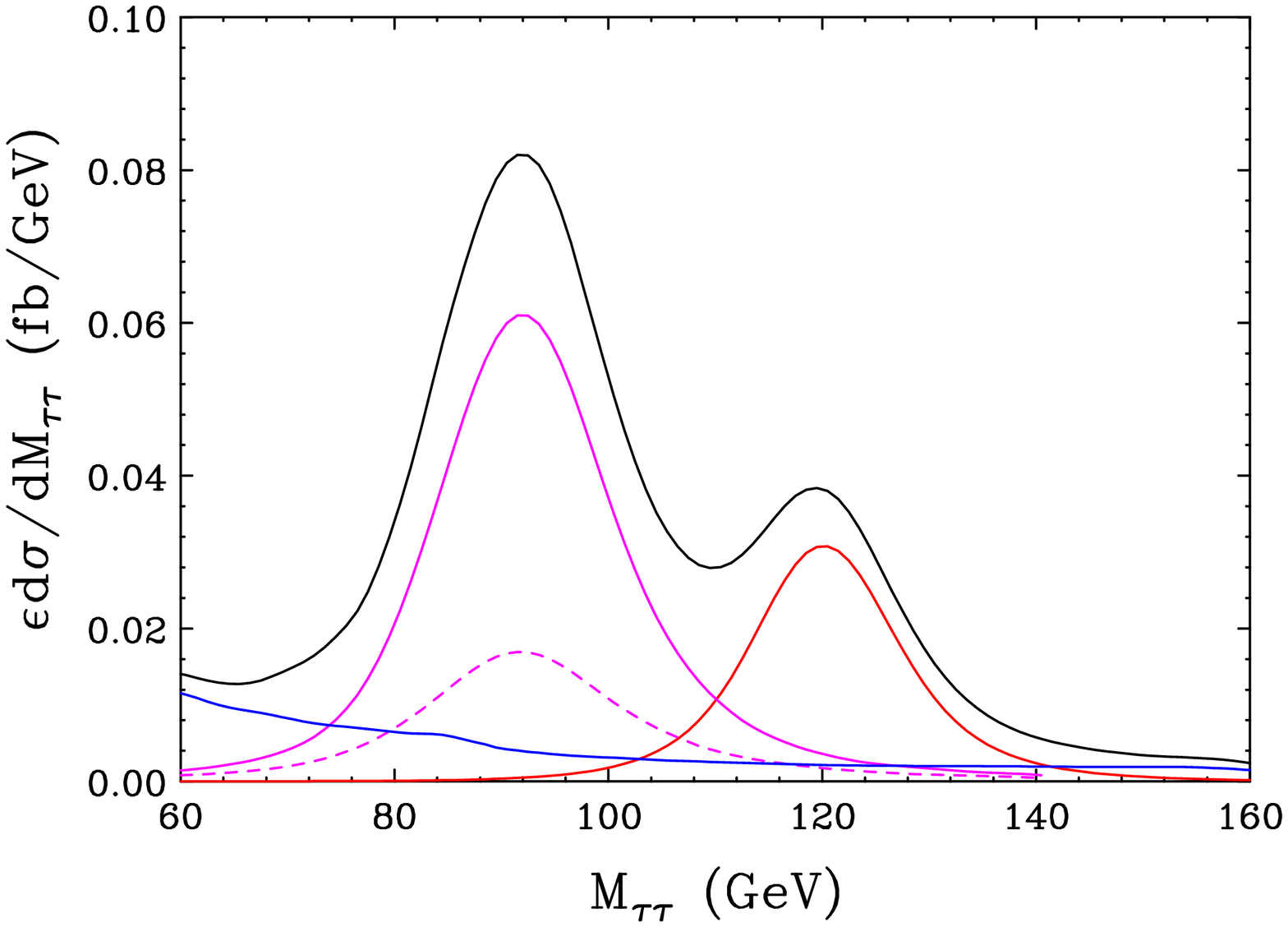}
\end{center} \vspace*{-0.2cm}
\caption{Reconstructed $\tau$ pair invariant mass distribution for a
SM $M_H=120$~GeV signal and backgrounds after the cuts of 
Eqs.~(\ref{eq:basic})-(\ref{eq:xtau}),(\ref{eq:ang}) and multiplication of 
the Monte Carlo results by the expected particle ID efficiencies and minijet 
veto survival probabilities. The double-peaked solid black line represents the 
sum of the signal and all backgrounds. Individual components are:
the $Hjj$ signal (red), the irreducible QCD $Zjj$ background (solid magenta), 
the irreducible EW $Zjj$ background (dashed magenta), and the
combined reducible backgrounds from QCD + EW + Higgs $WWjj$ events and 
$t\bar{t} + jets$ and $b\bar{b}jj$ production (blue).}
\label{fig:Mtautau}
\end{figure*}

The results summarized in Table~\ref{summary} show that it is possible to 
isolate a low-background $qq\to qqH,\;H\to\tau\tau\to e^\pm\mu^\mp\sla p_T$ 
signal for the SM Higgs boson at the LHC. Counting rates will be
sufficiently large to obtain a better than 
$5\sigma$ signal with $\approx 60$~fb$^{-1}$ of data for the mass range 
105-135~GeV. Extending the observability region down to 100~GeV still 
requires less than 90~fb$^{-1}$ of data, assuming low luminosity conditions.
Above $M_H=140$~GeV the signal quickly deteriorates, due 
to the falling branching ratio for the $H\to\tau\tau$ mode.
These results are comparable to those for the 
$H\to\tau\tau \to h^\pm \ell^\mp \sla{p}_T$ channel~\cite{thesis,RZ_tautau}, 
except as $M_H$ approaches 145-150 GeV. In this high mass range the purely
leptonic $\tau\tau$ decay signal receives substantial backgrounds from
$H\to WW$. Combined with the semileptonic channel discussed in 
Refs.~\cite{thesis,RZ_tautau} one effectively doubles the 
available statistics for the $H\to\tau\tau$ decay mode, making observation 
of this decay possible with significantly less than 60~fb$^{-1}$ of data, and 
ultimately providing a number of cross checks for the individual analyses. 
In the MSSM framework the combination of the two decay modes yields a 
$5\sigma$ signal for $\lesssim 40$~fb$^{-1}$ of data with an arbitrary choice 
of MSSM parameters which are still allowed by LEP data. 

The expected purity of the signal is demonstrated in Fig.~\ref{fig:Mtautau}, 
where the reconstructed $\tau\tau$ invariant mass distribution for a SM 
Higgs boson of mass 120~GeV is shown, together with the various backgrounds, 
after application of all cuts, ID efficiencies and a minijet veto. This 
purity is made possible because the weak boson fusion process, together 
with the $H\to\tau\tau \to e^\pm \mu^\mp \sla{p}_T$ decay, provides 
a complex signal, with a multitude of characteristics which distinguish it 
from the various backgrounds.

The basic feature of the $qq\to qqH$ signal is the presence of two forward 
tagging jets inside the acceptance region of the LHC detectors, of sizable 
$p_T$, and of dijet invariant mass in the TeV range. Typical QCD backgrounds, 
with isolated charged leptons and two hard jets, are much softer. 
In addition, the QCD backgrounds are dominated by $W$ bremsstrahlung off 
forward scattered quarks, which give typically higher-rapidity charged 
leptons. In contrast, the EW processes give rise to quite central leptons, 
and this includes not only the Higgs signal but also EW $WWjj$ and 
$\tau\tau jj$ production, which also proceed via weak boson fusion. It is 
this similarity that prevents one from ignoring EW analogs to background 
QCD processes, which a priori are smaller by two orders of magnitude in 
total cross section, but after basic cuts remain the same size as their 
QCD counterparts.

In addition to various invariant mass and angular cuts, one can discriminate
between the real $\tau$'s of the signal (and of the QCD and EW $\tau\tau jj$ 
backgrounds) and ``fake'' $\tau$'s from the $W,t,b$ backgrounds. 
This is possible because 
the high energy of the produced $\tau$'s makes their decay products almost 
collinear. Combined with the substantial $p_T$ of the $\tau^+\tau^-$ system 
this allows for $\tau$-pair mass reconstruction. The $W$ decays do not exhibit 
this collinearity due to their modest boost in the lab frame. This leads
to markedly different angular correlations 
between the $\sla{p_T}$ vector and the charged lepton momenta. 
Our real-$\tau$ criteria make use of these 
differences and largely eliminate the non-$\tau$ backgrounds.

We advocate taking advantage of an additional fundamental characteristic of 
QCD and EW processes. Color-singlet exchange in the $t$-channel, as 
encountered in Higgs boson production by weak boson fusion (and in the EW 
$Zjj$ background), leads to additional soft jet activity which differs 
strikingly from that expected for the QCD backgrounds in both geometry and 
hardness: gluon radiation in QCD processes is typically both harder and more 
central than in WBF processes. We exploit this radiation, via a veto on 
events with central minijets of $p_T > 20$~GeV, and expect a typical $70\%$ 
reduction in QCD backgrounds and about a $25\%$ suppression of EW backgrounds, 
but only about a $10\%$ loss of the signal.

We have identified the most important distributions for enhancing the signal 
relative to the background, and set the various cuts conservatively to avoid 
bias for a certain Higgs boson mass range. There is ample room for 
improvement of our results. A multivariate analysis of a complete set of 
signal and background distributions is expected to lead to improved 
background suppression. Mass specific cuts should eventually be employed 
and will improve matters as is evident from, e.g., the angular and lepton 
invariant mass distributions of Figs.~\ref{fig:angdist},\ref{fig:mll}. 
Additional suppression of the $t\bar{t} + jets$ 
background may be possible with $b$ identification and veto in the 
$p_{T_b} < 20$~GeV region. We do not pursue these questions here. One 
reason is that our results are derived at the parton level only. Even though
we have included expected detector resolution effects and losses due to finite
trigger and detection efficiencies, a more complete detector simulation is
now needed. We have to leave this work to our experimental colleagues. 

The very promising results of this study suggest that the 
$H\to\tau\tau \to ee,\mu\mu+\sla p_T$ modes should also be considered. 
The dilepton invariant mass distribution of Fig.~\ref{fig:mll} shows that
elimination of the $Z$ peak in $Z\to ee,\mu\mu$ backgrounds would reduce 
the Higgs signal by a small amount only. In addition, the requirement of 
significant $\sla p_T$ (Eq.~(\ref{eq:ptmiss})) is expected to largely 
eliminate QCD or EW $Zjj$ production, leaving $ZZjj$ and $ZWjj$ events
with invisible $Z$ or $W$ decays as the additional backgrounds. Given our
results for the analogous $WWjj$ events we expect these new backgrounds to 
be minimal. This implies that the purely leptonic $H\to\tau\tau$ signal
can most likely be enhanced by almost another factor of two, further reducing
the integrated luminosity required for observation of the 
$H\to\tau\tau$ signal.

Measuring the Higgs-fermion coupling will be an important test of the 
Standard Model as well as its supersymmetric extension. 
For such a measurement, 
via the analysis outlined in this paper, minijet veto probabilities must be 
precisely known. 
For calibration purposes, one can analyze $Zjj$ events at the LHC. The 
production rates of the QCD and EW $Zjj$ events can be reliably predicted 
and, thus, the observation of the $Z\to\ell\ell$ peak allows for a direct 
experimental assessment of the minijet veto efficiencies, in a kinematic 
configuration very similar to the Higgs signal.

{\it In Summary:} 
Observation of the SM or MSSM Higgs scalar(s) via 
$h/H\to\tau\tau \to e^\pm \mu^\mp \sla{p_T}$ in weak boson fusion is possible
at the LHC with modest integrated luminosities, if the Higgs boson lies 
in the mass range between about 100 and 140 GeV. Extending the search range
upward to 150 GeV should eventually be possible. 
Weak boson fusion at the LHC promises to be an exciting and important 
channel, both for validating the Standard Model via direct measurement of 
a Higgs-fermion coupling and as a low-luminosity ``see-or-die'' test of 
the MSSM.


\acknowledgements
We would like to thank R.~Kinnunen, A.~Nikitenko, E.~Richter-W\c{a}s 
and W.~Smith for
providing information on expected LHC detector performance.
This research was supported in part by the University of Wisconsin Research
Committee with funds granted by the Wisconsin Alumni Research Foundation and
in part by the U.~S.~Department of Energy under Contract
No.~DE-FG02-95ER40896. Fermilab is operated by URA under DOE contract 
No.~DE-AC02-76CH03000.


\end{document}